\documentclass[epj,twocolumn]{svjour_mod}
\usepackage{amsmath}
\usepackage{amssymb}
\usepackage{graphicx}
\usepackage[mathscr]{eucal}
\usepackage{xcolor}
\usepackage[english]{babel}
\usepackage{booktabs}
\usepackage{xspace}
\usepackage{multirow}
\usepackage{cite}

\newcommand\as{\alpha_s}
\newcommand{\beq}{\begin{equation}}
\newcommand{\eeq}{\end{equation}}
\newcommand\ord{{\cal O}}
\newcommand\clH{{\mathbb H}}
\newcommand\clS{{\mathbb S}}
\newcommand\aNLO{MadGraph5\_aMC@NLO\xspace}
\newcommand\aNLOs{MG5\_aMC\xspace}
\newcommand\aNLOD{MC@NLO-$\Delta$\xspace}
\newcommand\PYe{{\sc Pythia8}\xspace}
\newcommand\HWs{{\sc Herwig7}\xspace}
\newcommand\PhiB{\Phi_B}
\newcommand\Phir{\Phi_r}
\newcommand\confS{K_{\clS}}
\newcommand\confH{K_{\clH}}
\newcommand\GenMC{{\cal F}_{\rm MC}}
\newcommand\GenNLO{{\cal F}_{\rm MC@NLO}}
\newcommand\KMC{K^{(\rm MC)}}
\newcommand\KMCz{K^{(\rm MC,0)}}
\newcommand\sigmaH{\sigma^{(\clH)}}
\newcommand\sigmaS{\sigma^{(\clS)}}
\newcommand\muref{\mu_{\rm ref}}
\newcommand\Qkin{Q^{\rm kin}}
\newcommand\Qcut{Q_{\rm cut}}
\newcommand\Sfun{{\cal S}}
\newcommand\DZ{\Theta^{\rm DZ}}
\newcommand\pT{p_{T}}
\newcommand\HT{H_{T}}
\newcommand\SCALUP{{\tt SCALUP}\xspace}
\newcommand\param[1]{{\tt #1}}
\newcommand\bt{\bar{t}}

\newcommand\vnone{\texttt{none}\xspace}
\newcommand\vsumkl{\texttt{sum-kl}\xspace}
\newcommand\vsumij{\texttt{sum-ij}\xspace}
\newcommand\vmcflow{\texttt{mc-flow}\xspace}
\newcommand\vbs{\texttt{bornspr}\xspace}
\newcommand\vdelta{\texttt{delta}\xspace}
\newcommand\vall{\texttt{all}\xspace}

\usepackage[colorlinks=true,allcolors={blue!70!black}]{hyperref}
\usepackage{footnotebackref}

\begin{document}

\author{
Rikkert Frederix \inst{1}
\and
Paolo Torrielli \inst{2}
}
\date{}
\title{Born spreading with consistent shower starting scales in MC@NLO and FxFx merging}
\titlerunning{Born spreading with consistent shower starting scales}
\authorrunning{R.~Frederix, P.~Torrielli}
\institute{
Department of Physics, Lund University, Box 118, 221 00 Lund, Sweden
\and
Dipartimento di Fisica, Universit\`a di Torino and INFN, Sezione di Torino,
I-10125 Torino, Italy
\\
\email{rikkert.frederix@fysik.lu.se, paolo.torrielli@unito.it}
}

\abstract{
Negative weights are a major cost in NLO event generation matched to parton showers at the LHC, as they increase the number of events needed for a given statistical precision.
Born spreading reduces the fraction of negative-weight $\mathbb S$ events
in MC@NLO by redistributing the Born contribution over the radiation
variables. It introduces no bias only if everything that is passed to the
parton shower together with $\mathbb S$ events depends on Born-level
quantities alone. We show that the same condition is needed for the MC
counterterms to reproduce the first shower emission, independently of how
the radiation phase space is integrated. In \aNLO, the $\mathbb S$-event
starting scales did not fully satisfy this condition, which so far prevented the
public release of Born spreading. We present a new and consistent
construction of the starting scales for ordinary MC@NLO, FxFx merging and
\aNLOD, with which Born spreading is now publicly available, including
for merged samples. We also study modified formulations of the
MC counterterms, and FxFx-merged $W$+jets production with up to two jets
at NLO, for which Born spreading reduces the fraction of negative weights
after merging without affecting the predictions, and the event samples
are generated six times faster than with the former implementation.
}
\maketitle

\section{Introduction}
\label{sec:intro}

Next-to-leading order (NLO) QCD calculations matched to parton-shower
event generators (NLO+PS) are a standard tool of LHC phenomenology, and
account for the bulk of the simulations employed by the experimental
collaborations, whose computing needs are steadily
increasing~\cite{HEPSoftwareFoundation:2020daq,CERN-LHCC-2022-005,Software:2815292}.
Among the matching methods, the most widely used are
MC@NLO~\cite{Frixione:2002ik} and
POWHEG~\cite{Nason:2004rx,Frixione:2007vw}, while several alternatives
have been proposed, such as KrkNLO~\cite{Jadach:2015mza}, MAcNLOPS~\cite{Nason:2021xke}
and ESME~\cite{vanBeekveld:2025lpz}. A practical drawback of MC@NLO is
the presence of events with negative weights: with a fraction $f$ of such
events, the number of events needed for a given statistical precision
grows as $(1-2f)^{-2}$, and so does the cost of their shower and detector
simulation. Several techniques have therefore been devised to reduce $f$,
such as
folding~\cite{Nason:2007vt,Frixione:2007nu,Alioli:2010xd,Frederix:2020trv},
\aNLOD~\cite{Frederix:2020trv}, Born
spreading~\cite{Frederix:2023hom,Che:2024bsc}, and the reweighting of
Born-level events~\cite{ElFarkh:2026wtx}, while negative weights can also
be removed from existing event samples by cell
resampling~\cite{Andersen:2020sjs,Andersen:2021mvw,Andersen:2023cku,Andersen:2024mqh}. Born spreading targets the
negative weights of the $\clS$ events, i.e.~of the events with Born-level
kinematics: since the Born contribution does not depend on the radiation
variables, it can be redistributed over them, and concentrated where the
rest of the $\clS$-event integrand is negative, at a small, fixed
computational cost. This leaves the predictions unchanged, provided that
the $\clS$ events are blind to the radiation variables.

This proviso turns out to be a statement about the shower starting
scales. In MC@NLO, the double counting between the real-emission matrix
elements and the first emission of the shower is removed by means of the
MC counterterms, namely the ${\cal O}(\as)$ expansion of the shower acting
on Born-level configurations. This makes the matching sensitive to all of
the features of the shower which affect its first emission: its
evolution variable, splitting kernels, kinematic maps, dead zones and
colour assignments, and its {\em starting scale}, i.e.~the upper bound on
the hardness of the emissions that the shower can generate from a given
hard event. The choice of the starting scale is largely arbitrary, subject
only to the requirement that it be of the order of the hard scale of the
process; however, once this choice has been made, the MC
counterterms must reflect it exactly. In the automated implementation of
MC@NLO in \aNLO~\cite{Alwall:2014hca} (\aNLOs henceforth), which is based
on the FKS subtraction~\cite{Frixione:1995ms,Frixione:1997np,Frederix:2009yq},
the starting scale is randomised by means of a smooth damping function,
which multiplies the MC counterterms~\cite{Frederix:2012ps,Alwall:2014hca}.
The randomisation avoids sharp features in the region of intermediate
hardness, where the matrix-element and shower descriptions are joined.
Although the $\clS$ events have Born-level kinematics, their
short-distance cross section is integrated over the full real-emission
phase space. As we shall show, MC@NLO is then consistent, whatever the way
in which this integration is performed, only if everything that is passed
to the shower together with an $\clS$ event depends on Born-level
quantities alone. This is precisely the condition required by Born
spreading.

Over the years, the assignment of the shower starting scales in \aNLOs
has evolved through a sequence of extensions (to different parton
showers, to FxFx merging~\cite{Frederix:2012ps,Frederix:2021agh}, to
\aNLOD~\cite{Frederix:2020trv}), and it has come to include several
technical lower bounds on the scales, as well as dependences of the
$\clS$-event starting scale on the sampled real-emission configuration,
most notably on whether the latter lies in a dead zone of the shower.
Such dependences are not compatible with Born spreading, which could
therefore not be included in the public version of the code.

In this paper we present a new treatment of the shower starting scales
for MC@NLO, implemented in \aNLOs as part of a complete restructuring of
its MC counterterms, with which Born spreading has become publicly
available, for the first time also in conjunction with FxFx merging. We
focus on \PYe~\cite{Bierlich:2022pfr}, and consider ordinary MC@NLO, its
use in FxFx merging, and \aNLOD. The new construction is formulated in
terms of Born-level quantities only, with a single reference scale and a
single random number per event, and the kinematic limits of the shower
enter the MC counterterms exactly as they enter the shower. Besides
solving the problems mentioned above, it decouples the starting scales
from the FKS sectors, which has allowed us to investigate several
modified formulations of the interplay between MC counterterms, FKS
sectors and colour flows, which also affect the fraction of negative
weights.

These developments are particularly relevant for multi-jet merged
samples, which are among the most CPU-intensive ones in LHC
simulations~\cite{HEPSoftwareFoundation:2020daq,CERN-LHCC-2022-005,Software:2815292}.
FxFx-merged samples, in particular for vector-boson plus jets
production, are routinely used by the experimental collaborations, both
in measurements and in searches, see
e.g.~Refs.~\cite{ATLAS:2024nrd,ATLAS:2024mow,ATLAS:2026wkp,ATLAS:2024moy,CMS:2026fvm,CMS:2026yrb},
and for them both the CPU time spent in
the merging and the fraction of negative weights, which is large for
processes with light partons at the Born level~\cite{Frederix:2023hom},
are a concern. Besides illustrating the new construction for single
processes, we therefore study in detail FxFx-merged $W$+jets production
with up to two jets at NLO, with which we quantify the merging
efficiency, the fraction of negative weights before and after merging,
the effect of Born spreading, the CPU time, and the impact of the new
construction on physical distributions.

The paper is organised as follows. In Section~\ref{sec:cons} we derive the
consistency condition that links the starting scales with the MC
counterterms. In Section~\ref{sec:old} we review the scale assignment
used so far in \aNLOs, and discuss where it departs from such a condition. The
new construction is presented in Section~\ref{sec:new}, for ordinary
MC@NLO, FxFx and \aNLOD. In Section~\ref{sec:opp} we describe the
public implementation of Born spreading and the modified formulations of
the matching that the new construction has made possible. In
Section~\ref{sec:res} we present results for the fraction of negative
weights and for physical distributions for four processes, and in
Section~\ref{sec:resFxFx} for FxFx-merged $W$+jets production. We draw
our conclusions and give an outlook in Section~\ref{sec:conc}.

\section{Starting scales and MC counterterms}
\label{sec:cons}

\subsection{The MC@NLO cross section}
\label{sec:mcnlo}
We keep the notation to a minimum, and follow
Refs.~\cite{Frederix:2020trv,Frederix:2023hom}. The formulae below
refer to a given real-emission process, FKS parton $i$, and sister $j$,
the sums over these being understood. We denote by $\PhiB$ the Born-level
integration variables (including the Bjorken $x$'s), and by
$\Phir=\{\xi,y,\varphi\}$ the FKS radiation variables; $\xi\to 0$ and
$y\to 1$ correspond to the soft and collinear limits, respectively. The
Born ($\confS$) and real-emission ($\confH$) kinematic configurations
are such that $\confS=\confS(\PhiB)$, $\confH=\confH(\PhiB,\Phir)$, and
$\confH\to\confS$ in the soft and collinear limits~\cite{Frederix:2009yq}.
The MC@NLO generating functional reads
\beq
\GenNLO=\GenMC\!\left(\confH;Q_\clH\right)d\sigmaH
+\GenMC\!\left(\confS;Q,c\right)d\sigmaS\,,
\label{eq:genMCatNLO}
\eeq
where $\GenMC$ is the generating functional of the parton shower, for
which we have made explicit the dependence on the starting scale(s)
($Q_\clH$ and $Q$ for $\clH$ and $\clS$ events, respectively) and, for
$\clS$ events, on the colour flow $c$, which defines the colour dipoles
the shower radiates from. The short-distance cross sections are
\begin{eqnarray}
d\sigmaH&=&\Big[R-\KMC\Big]\,d\PhiB\,d\Phir\,,
\label{eq:xsecH}
\\
d\sigmaS&=&\Big[\frac{B+V}{\int d\Phir}+\KMC-C\Big]\,d\PhiB\,d\Phir\,,
\label{eq:xsecS}
\end{eqnarray}
with $R$ the real-emission matrix element, $C$ the FKS soft, collinear
and soft-collinear counterterms, $B$ the Born, $V$ the finite part of the
virtual corrections (including the integrated FKS counterterms), and
$\KMC$ the MC counterterms. Eq.~\eqref{eq:xsecS} makes it clear that,
although $\clS$ events have Born kinematics, their short-distance cross
section has support in the whole $(\PhiB,\Phir)$ space.

The parton showers we are interested in work at the leading-colour
level: the first emission off a Born configuration is generated from the
colour dipoles of a single leading colour flow $c$, which is not chosen
by the shower, but read from the event file. The MC counterterms must
therefore assume definite probabilities for the flows; we use
$\pi_c=|{\cal A}_c|^2/\sum_{c'}|{\cal A}_{c'}|^2$, with ${\cal A}_c$ the
leading-colour amplitude associated with flow $c$. The undamped MC
counterterms read
\begin{multline}
\KMCz(\PhiB,\Phir)=B(\PhiB)\sum_c\pi_c
\\\times
\sum_{b\in c}k_{mb}(\PhiB,\Phir)\,\DZ_{mb}(\PhiB,\Phir)\,,
\label{eq:KMC0}
\end{multline}
where $m$ is the Born parton that branches into $i$ and $j$, the inner sum
runs over its colour partner(s) $b$ in flow $c$, $k_{mb}$ is the
${\cal O}(\as)$ branching density of the shower for an emission off the
$(m,b)$ dipole, and $\DZ_{mb}$ is equal to one where the shower can
generate the configuration $\confH(\PhiB,\Phir)$ and to zero in its dead
zones~\cite{Frixione:2002ik,Alwall:2014hca}. In \aNLOs, the counterterms
are damped as follows:
\beq
\KMC(\PhiB,\Phir)=D\big(q(\PhiB,\Phir);\mu_1,\mu_2\big)\,
\KMCz(\PhiB,\Phir)\,,
\label{eq:KMCdamp}
\eeq
where $q$ is the hardness of the emission in the shower language (the
relative transverse momentum that serves as evolution variable in the
case of \PYe), and
\beq
D(q;\mu_1,\mu_2)=\left\{
\begin{array}{ll}
1 & q\le\mu_1\,,\\
1-F\left(\frac{q-\mu_1}{\mu_2-\mu_1}\right) & \mu_1<q\le\mu_2\,,\\
0 & q>\mu_2\,,
\end{array}
\right.
\label{eq:Ddef}
\eeq
with $F(x)=x^2/(x^2+(1-x)^2)$. The two parameters
$\mu_1\le\mu_2$ are of the order of the hardness of the process,
$\mu_\alpha=f_\alpha\muref$ with $f_1=0.1$ and $f_2=1$ by default, and
$\muref$ a reference scale. In the soft limit, where the shower does not
reproduce the eikonal behaviour of $R$, $\KMC$ is smoothly replaced by the
soft FKS counterterm, by means of the function ${\cal
G}$~\cite{Frixione:2002ik,Alwall:2014hca}; we shall come back to this
point in Section~\ref{sec:newKMC}.

\subsection{The consistency condition}
\label{sec:cond}
As we now show, the damping function $D$ must be the probability that
the phase space available to the shower extends beyond hardness $q$,
and it thus determines the way in which the starting scales must be
chosen\footnote{We thank Stefano Frixione for extensive discussions on this point.}. Since this is the central issue, we examine it in detail.
Suppose that the starting scale $Q$ and the colour flow $c$ that are
passed to the shower together with an $\clS$ event stemming from the
integration point $(\PhiB,\Phir)$ are chosen with probability density
$P(Q,c|\PhiB,\Phir)$. This density is a property of the event generation, not of the shower, and its dependence on $c$ need not coincide with $\pi_c$:
for example, the flows may be restricted to those compatible with the
integration channel, or chosen according to quantities computed at
$(\PhiB,\Phir)$ (see Section~\ref{sec:old}).
At ${\cal O}(\as)$, the shower generating
functional is
\begin{multline}
\GenMC\!\left(\confS;Q,c\right)={\cal F}_0(\confS)
+\sum_{b\in c}\int d\Phir^\prime\,k_{mb}\,\DZ_{mb}
\\\times
\Theta\big(Q-q^\prime\big)
\Big[{\cal F}_0(\confH^\prime)-{\cal F}_0(\confS)\Big]+\ord(\as^2)\,,
\label{eq:GenMCexp}
\end{multline}
where ${\cal F}_0(K)$ is the trivial generating functional of the
configuration $K$, $q^\prime=q(\PhiB,\Phir^\prime)$ and
$\confH^\prime=\confH(\PhiB,\Phir^\prime)$; we have suppressed the
arguments of $k_{mb}$ and $\DZ_{mb}$, which are $(\PhiB,\Phir^\prime)$.
Note that the emission variables $\Phir^\prime$ that appear in
Eq.~\eqref{eq:GenMCexp} have nothing to do with the integration variables
$\Phir$ of Eq.~\eqref{eq:xsecS}, which are integrated over. At
${\cal O}(\as)$, only the Born term in Eq.~\eqref{eq:xsecS} contributes
to the emission term of Eq.~\eqref{eq:GenMCexp}, which gives
\begin{multline}
\int d\PhiB\,B\int d\Phir\,\rho(\Phir|\PhiB)\sum_c\int dQ\,
P(Q,c|\PhiB,\Phir)
\\\times
\sum_{b\in c}\int d\Phir^\prime\,k_{mb}\,\DZ_{mb}\,\Theta\big(Q-q^\prime\big)
\Big[{\cal F}_0(\confH^\prime)-{\cal F}_0(\confS)\Big]\,.
\label{eq:Sexp}
\end{multline}
Here, $\rho(\Phir|\PhiB)$, with $\int d\Phir\,\rho=1$, is the
distribution of the Born contribution over the radiation variables:
$\rho=1/\int d\Phir$ in Eq.~\eqref{eq:xsecS}, where the flat measure
$d\Phir$ is defined by the parametrisation of the radiation phase space
used in the integration, while $\rho$ is a non-trivial function with
Born spreading~\cite{Frederix:2023hom}, see Section~\ref{sec:BS}. By
defining the effective probabilities
\begin{multline}
{\cal P}_c(q|\PhiB)=\int d\Phir\,\rho(\Phir|\PhiB)
\\\times
\int dQ\,P(Q,c|\PhiB,\Phir)\,\Theta(Q-q)\,,
\label{eq:Peff}
\end{multline}
one sees that the shower generates, from the $\clS$ events, the
${\cal O}(\as)$ emission density
\mbox{$B\sum_c{\cal P}_c(q)\sum_{b\in c}k_{mb}\DZ_{mb}$}. In MC@NLO, this
must coincide with the MC counterterms, which are subtracted from the
$\clH$ events and added back to the $\clS$ events, so as to avoid double
counting~\cite{Frixione:2002ik}. By comparing this density with
Eqs.~\eqref{eq:KMC0} and~\eqref{eq:KMCdamp}, we obtain the consistency
condition
\beq
{\cal P}_c(q|\PhiB)=\pi_c(\PhiB)\,D\big(q;\mu_1(\PhiB),\mu_2(\PhiB)\big)\,,
\label{eq:cond}
\eeq
which must be fulfilled for all $q$, $c$ and $\PhiB$. Crucially, the
left-hand side of Eq.~\eqref{eq:cond} depends on $\rho$, whereas its
right-hand side does not. The distribution $\rho$ is a technical
quantity: it changes with the parametrisation of the radiation phase
space, and it is effectively modified by
folding~\cite{Nason:2007vt,Frixione:2007nu,Alioli:2010xd,Frederix:2020trv},
where the event inherits the information associated with one of
several $\Phir$ points, and by Born spreading. Equation~\eqref{eq:cond} can
be satisfied for any $\rho$ if and only if
\begin{eqnarray}
P(Q,c|\PhiB,\Phir)&=&\pi_c(\PhiB)\,P_D(Q|\PhiB)\,,
\nonumber
\\
\int_q^\infty dQ\,P_D(Q|\PhiB)&=&D(q;\mu_1,\mu_2)\,.
\label{eq:solution}
\end{eqnarray}
In other words, the colour flow passed to the shower must be generated
with the probabilities $\pi_c$ that are assumed in the MC counterterms,
Eq.~\eqref{eq:KMC0}, and the starting scale must be generated as
\beq
Q=D^{-1}(r;\mu_1,\mu_2)=\mu_1+(\mu_2-\mu_1)\,
\frac{\sqrt{r}}{\sqrt{r}+\sqrt{1-r}}\,,
\label{eq:Qdraw}
\eeq
with $r$ a flat random number, both {\em independently of the radiation
variables} $\Phir$. We stress that nothing that is passed to the shower
together with an $\clS$ event may depend on $\Phir$, although it may
depend on $\PhiB$ and on random numbers drawn independently of $\Phir$.
On the other hand, it is not necessary that the scale be drawn only
once per Born configuration: it can be generated anew in each FKS sector
or fold, the event inheriting the scale of the contribution which is
eventually selected (with a probability that depends on $\Phir$). This
is because selecting one of several independent draws of the same
distribution, by means of a criterion that does not depend on their
values, does not alter their distribution.

\subsection{Kinematic limits of the shower}
\label{sec:kin}
\begin{figure*}[t]
\begin{center}
\includegraphics[width=\textwidth]{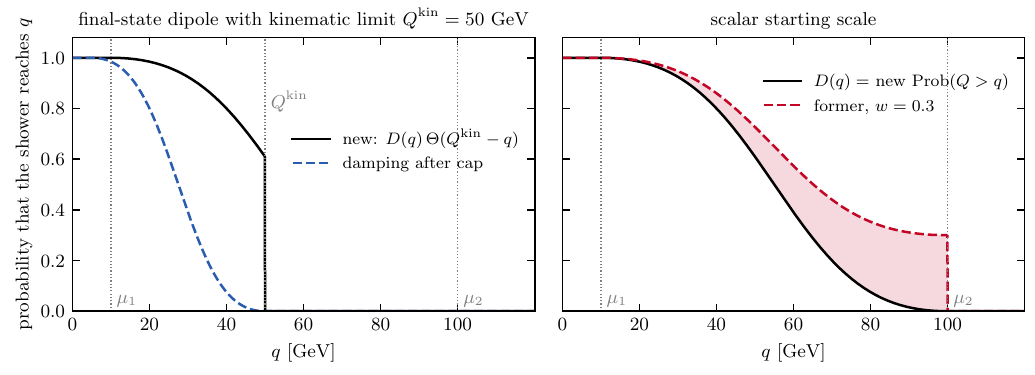}
\caption{Probability that the phase space available to the shower started
from an $\clS$ event extends beyond hardness $q$, for $\mu_1=10$~GeV and
$\mu_2=100$~GeV. Left panel: a final-state
dipole with kinematic limit $\Qkin=50$~GeV. With a single \SCALUP, the
shower phase space is bounded by $\min(Q,\Qkin)$ (solid), whereas
damping a reference scale capped by the kinematics would give the dashed curve. Right panel: a scalar starting scale (e.g.~an
initial-state emitter in \PYe). The new prescription (solid) coincides
with the damping $D(q)$ of the MC counterterms; in the former one (dashed),
the $\clS$ events whose integration point lies in the region where the
MC counterterms vanish, here a fraction $w=0.3$ of the Born contribution
(see Eq.~\eqref{eq:Pold}), are assigned the scale $\mu_2$. The shaded area
is not compensated by the MC counterterms.}
\label{fig:damping}
\end{center}
\end{figure*}

Besides the starting scale, \PYe imposes further, dipole-dependent,
kinematic limits on the phase space of the shower. In ordinary MC@NLO,
\PYe is given a single starting scale per event, i.e.~the \SCALUP
parameter of the Les Houches accord~\cite{Boos:2001cv,Alwall:2006yp}.
With the settings employed for MC@NLO
matching~\cite{Bierlich:2022pfr}, an emission off a final-state parton $a$
colour-connected with parton $b$ then has a transverse momentum smaller
than
\beq
Q_{ab}=\min\big(Q,\Qkin_{ab}\big)\,,
\;\;\;\;\;\;
\Qkin_{ab}=\frac{1}{2}\sqrt{\big(M_{ab}-m_b\big)^2-m_a^2}\,,
\label{eq:Qab}
\eeq
where $M_{ab}$ is the invariant mass of the dipole and $m_a$, $m_b$
are the masses of its ends, while for an initial-state parton $a$ one has
$Q_{ab}=Q$. It is also possible to give \PYe independent starting scales
for each colour dipole, by means of dipole-tagged scales in the event
file~\cite{Frederix:2020trv}, to which the kinematic limits are still
applied. This is what is done in \aNLOD, where the scales $Q_{ab}$ of
Eq.~\eqref{eq:Qab} are themselves written in the event file
(Section~\ref{sec:newDelta}). In either case, the shower phase space is
bounded by \mbox{$q<\min(Q,\Qkin_{ab})$},
 and Eq.~\eqref{eq:cond} 
becomes
\beq
{\cal P}_{c,mb}(q|\PhiB)=\pi_c\,D(q;\mu_1,\mu_2)\,\Theta\big(\Qkin_{mb}-q\big)\,.
\label{eq:condkin}
\eeq
That is, the MC counterterms must be damped by means of the distribution
of the scale that is actually written in the event file, and
independently cut by the kinematic limits. Note that Eq.~\eqref{eq:condkin}
is not equivalent to damping a reference scale capped by the
kinematics, i.e.~to the function
\mbox{$D(q;f_1\min(\muref,\Qkin_{mb}),f_2\min(\muref,\Qkin_{mb}))$}.
In particular, with a single \SCALUP
there is a finite probability, equal to $D(\Qkin_{mb})$, that the shower
phase space extends up to the kinematic endpoint. This is illustrated in
the left panel of Fig.~\ref{fig:damping}.

\subsection{$\clH$ events}
\label{sec:Hcons}
For $\clH$ events the choice of the starting scale $Q_\clH$ is not
constrained by the ${\cal O}(\as)$ consistency of the matching: since the
$\clH$-event cross section is already of ${\cal O}(\as)$ relative to the
Born, a change of $Q_\clH$ affects the predictions only at
${\cal O}(\as^2)$. Nevertheless, some choices are more natural than
others. Soft and collinear $\clH$ events are largely cancelled by $\clS$
events, and it is therefore desirable that in the soft and collinear
limits $Q_\clH$ approach the starting scale of the corresponding $\clS$
event, so that the two are showered in a similar way. Away from these
limits, an $\clH$ event is a genuine hard configuration with one more
parton, and, as for the $\clS$ events, $Q_\clH$ is best chosen of the
order of the hardness of $\confH$, rather than of a scale associated with
the underlying Born configuration only.

\section{The former implementation}
\label{sec:old}
We now describe the assignment of the starting scales in the current
releases of \aNLOs (up to version 3.8); for definiteness, we refer to
matching with \PYe. In ordinary MC@NLO, the $\clS$-event starting scale
results from the following steps.
\begin{itemize}
\item[{\em a)}] The reference scale is $\muref=\HT/2$, i.e.~half the sum of
the transverse masses of all of the Born-level final-state particles,
bounded from below by $30$~GeV.
\item[{\em b)}] The damping interval is
\mbox{$\mu_1=\max(f f_1\muref,10\;{\rm GeV})$},
\mbox{$\mu_2=\max(f f_2\muref,\mu_1+20\;{\rm GeV})$}, with
$\mu_2\le\sqrt{S}$, and $f$ the user-defined shower-scale factor; if
\mbox{$\mu_2-\mu_1<10^{-3}\mu_2$}, $D$ is replaced by a $\Theta$ function. A
scale $Q_D$ is generated as in Eq.~\eqref{eq:Qdraw}.
\item[{\em c)}] A second reference scale, $Q_B$, is defined as the
smallest $k_T$-clustering scale~\cite{Catani:1993hr} of the light Born-level
partons, or as the partonic c.m.~energy for processes without light
partons at the Born level. The event is assigned
\mbox{$\SCALUP=\max(\min(Q_D,Q_B),3\;{\rm GeV})$}, while the MC
counterterms are set to zero for
\mbox{$q>\max(\min(\mu_2,Q_B,\Qkin_{mb}),3\;{\rm GeV})$}.
\item[{\em d)}] If, at the sampled integration point $(\PhiB,\Phir)$,
the hardness $q$ exceeds $\mu_2$, or the configuration $\confH$ lies in a
dead zone for all of the colour partners of $m$ (i.e.~whenever $\KMC=0$
there), $Q_D$ is replaced by $\mu_2$.
\item[{\em e)}] If the smallest $k_T$-clustering scale of $\confH(\PhiB,\Phir)$
exceeds $Q_B$, the MC counterterms are set to zero, and the event is
assigned $\SCALUP=Q_B$, without any damping.
\item[{\em f)}] The colour flow is chosen when the event is written, with
probabilities proportional to $|{\cal A}_c|^2$, restricted to the flows
which are compatible with the Born diagram that defines the multi-channel
integration channel.
\end{itemize}
For $\clH$ events, \mbox{$\SCALUP=\min(\mu_2,\sqrt{\hat{s}}-\pT)$} for
processes without light partons at the Born level, with $\hat{s}$ the
partonic c.m.~energy squared and $\pT$ the transverse momentum of the
FKS parton, and
\mbox{$\SCALUP=\min(\mu_2,\max(d_n-d_{n+1}/2,d_n-\min(Q_D,\mu_2)))$}
otherwise, with $d_n$ and $d_{n+1}$ the two smallest $k_T$-clustering
scales of $\confH$.

In \aNLOD~\cite{Frederix:2020trv}, dipole-specific starting scales
$\mu_{ab}$ were used, each one generated independently with
Eq.~\eqref{eq:Qdraw}, with reference scale
\mbox{$\muref=\min(M_{ab},X)$}, where $X$ is the partonic c.m.~energy,
lowered to the smallest transverse mass of the massive coloured particles
and to the smallest $k_T$-clustering scale of the light partons. The
replacement of item {\em d)} was applied dipole by dipole, by using the
dipole-specific hardness of the sampled real emission. Furthermore, the
colour flow was chosen with probabilities proportional to the
flow-resolved MC counterterms at the sampled point $(\PhiB,\Phir)$
(rule {\em R.2} of Ref.~\cite{Frederix:2020trv}). In FxFx
merging~\cite{Frederix:2012ps}, the scale $Q_B$ was replaced by the
smallest clustering scale $d_{\min}(\confS)$ of the merging algorithm,
which is also the lowest scale of the Sudakov factors that reweight the
short-distance cross sections, while all of the other elements were left
unchanged.

Items {\em d)} and {\em e)}, and the choice of the colour flow in \aNLOD,
introduce a dependence on $\Phir$, and thus violate
Eq.~\eqref{eq:solution}. The consequences are easily worked out for
item {\em d)}. Denote by ${\cal R}_0(\PhiB)$ the region of the radiation
phase space where $\KMC=0$ at given $\PhiB$, and by
\mbox{$w(\PhiB)=\int_{{\cal R}_0}d\Phir\,\rho$} the fraction of the
Born contribution associated with it. From Eq.~\eqref{eq:Peff} one obtains
\beq
{\cal P}(q|\PhiB)=D(q)+w(\PhiB)\big[1-D(q)\big]\,\Theta(\mu_2-q)\,,
\label{eq:Pold}
\eeq
which is shown in the right panel of Fig.~\ref{fig:damping} for $w=0.3$.
Therefore, the shower generates from the $\clS$ events the additional
${\cal O}(\as)$ emission density $Bw(1-D)k_{mb}\DZ_{mb}$ in the range
$\mu_1<q<\mu_2$, which is not compensated by the MC counterterms. Thus,
strictly speaking, double counting occurs at ${\cal O}(\as)$ in the
region of intermediate hardness, and the formal NLO accuracy of the
predictions is lost there. The size of this effect depends on the
process and on the observable, but it is not under control, since $w$
depends on the parametrisation of the radiation phase space, and is
modified by folding and by Born spreading. Analogous considerations
apply to item {\em e)}, which in addition modifies the MC counterterms in
a region defined by a quantity, the $k_T$-clustering scale, unrelated to
the shower, and to the colour-flow assignment of \aNLOD.

The other features of the former implementation are not inconsistent
per se, but they are ad hoc and lead to unphysical behaviours. The
damping interval and the shower phase space of the MC counterterms are
defined in terms of two different reference scales, $\HT/2$ and $Q_B$,
which are combined by taking their minimum. The lower bounds of items
{\em a)} and {\em b)} completely determine the scales of processes
whose hardness is a few tens of GeV. The $\clH$-event scale for
processes without light Born-level partons does not approach the
starting scale of the $\clS$ events in the soft and collinear limits:
for $\pT\to 0$, the former tends to $\mu_2$, while the latter is
distributed between $\mu_1$ and $\mu_2$. In item {\em f)}, the
colour flows are restricted to those compatible with the integration
channel, so that their probabilities are not those that enter the MC
counterterms, i.e.~$\pi_c$. Finally, in
FxFx the $\clS$-event starting scale $\min(Q_D,d_{\min})$ is often
smaller than $d_{\min}(\confS)$: in that case, emissions with a hardness
between these two scales are Sudakov-suppressed neither by the FxFx
reweighting nor by the shower.

The $\Phir$ dependence of the $\clS$-event starting scales is the reason
why Born spreading could not be released. With Born spreading, $\rho$ is
deliberately concentrated in the regions where the rest of the $\clS$
integrand is negative, which changes $w$ in Eq.~\eqref{eq:Pold} and, in
turn, the distribution of the $\clS$-event starting scales. This bias,
which would not be a higher-order effect, was pointed out in
Ref.~\cite{Frederix:2023hom}, where it was avoided by a minimal
modification of the scale assigned to the $\clS$ events in the dead
zones. However, the scale assignment was entangled with all of the
other elements listed above (and with the FKS sectors, as we shall
discuss in Section~\ref{sec:opp}), and a consistent treatment
required its complete restructuring, which we now present.

\section{The new implementation}
\label{sec:new}
The new construction is based on the following principles:
\begin{itemize}
\item[{\em i)}] all of the information passed to the shower together with
$\clS$ events (colour flow, \SCALUP, dipole scales) is generated at the
Born level, before the radiation variables are sampled, and is shared by
all of the FKS sectors, folds, and $\clS$ and $\clH$ contributions
associated with the same Born configuration;
\item[{\em ii)}] a single reference scale is used;
\item[{\em iii)}] the MC counterterms are the exact ${\cal O}(\as)$
expansion of the shower started with these scales, including the
kinematic limits of the shower, according to Eq.~\eqref{eq:condkin};
\item[{\em iv)}] no bounds are imposed other than an infrared one.
\end{itemize}

\subsection{Born-level ingredients}
\label{sec:Bornlevel}
For each Born configuration, a leading colour flow $c$ is generated with
probability $\pi_c$, by considering all of the leading-colour flows. The
same flow is used for all of the FKS sectors and folds, for both $\clS$
and $\clH$ contributions, and it is the one written in the event file;
the colour flow of the $\clH$ events is obtained from $c$ by splitting
the colour line(s) of the mother parton $m$, as in
Ref.~\cite{Frederix:2020trv}. By default, the MC counterterms are
summed explicitly over the flows, with the weights $\pi_c$ of
Eq.~\eqref{eq:KMC0}. Alternatively, they can be evaluated at the sampled
flow only, i.e.~without the factor $\pi_c$ of Eq.~\eqref{eq:KMC0}, the
flow being sampled with that probability, so that the colour lines that
enter $\KMC$, its dead zones and its starting scales coincide event by
event with those that the shower will use (Section~\ref{sec:sectors}).

The reference scale $\muref(K)$ of a configuration $K$ is the smallest
scale in the clustering history of $K$ which is used in FxFx
merging~\cite{Frederix:2012ps}. This history is reconstructed by
following the Feynman-diagram topologies of the process, with a
$k_T$-type measure, in the spirit of the CKKW
procedure~\cite{Catani:2001cc}. For processes with light partons at the
Born level, $\muref$ is therefore the smallest of their clustering
scales, i.e.~the hardness of the softest jet; for processes without, it
is the hardness of the core process, e.g.~the invariant mass for the
production of a colour-singlet system. Thus, the same quantity defines
the $\clS$- and $\clH$-event scales (with $K=\confS$ and $K=\confH$,
respectively) and the lowest scale of the Sudakov reweighting in FxFx
merging.

\subsection{$\clS$-event starting scales}
\label{sec:newS}
The damping interval is
\begin{eqnarray}
\mu_1&=&\max\big(f f_1\muref(\confS),\Qcut\big)\,,
\nonumber
\\
\mu_2&=&\min\Big(\max\big(f f_2\muref(\confS),\mu_1+\delta\big),\sqrt{S}\Big)\,,
\label{eq:mu12}
\end{eqnarray}
where $f$ is the shower-scale factor (equal to one by default), and the
parameters $\Qcut=\delta=3$~GeV ensure that the starting scale lies above
the infrared cutoff of the shower, and that the damping function is not
arbitrarily steep. These are the only technical parameters left; in
particular, the $10$, $20$ and $30$~GeV bounds of Section~\ref{sec:old} have
been removed. A single scale $Q$ is generated with Eq.~\eqref{eq:Qdraw}
and written in the event file as \SCALUP. The effective dipole scales,
which are used internally in the construction of the MC counterterms and
of the $\clH$-event scales, and which are written in the event file in
the case of \aNLOD (see Section~\ref{sec:newDelta}), are given by
Eq.~\eqref{eq:Qab}. Here, $M_{ab}$ is computed with positive-energy
momenta also for an initial-state partner $b$, and on-shell masses are
used for $m_a$ and $m_b$, as is done in \PYe (version~8.313,
with the \param{TimeShower:limitPTmaxGlobal} option employed for MC@NLO
matching~\cite{Bierlich:2022pfr}).

\subsection{MC counterterms}
\label{sec:newKMC}
The MC counterterms read
\beq
\KMC=B\sum_{b\in c}k_{mb}\,\DZ_{mb}\,
\Theta\big(\Qkin_{mb}-q_{mb}\big)\,D(q_{mb};\mu_1,\mu_2)\,,
\label{eq:KMCnew}
\eeq
with $q_{mb}$ the \PYe evolution variable of the emission off the
$(m,b)$ dipole. Equation~\eqref{eq:KMCnew} refers to the sampled flow
$c$, as with the \vmcflow option of Section~\ref{sec:sectors}; by
default, it is summed over the leading colour flows with weights
$\pi_c$, as in Eq.~\eqref{eq:KMC0}. The
dead-zone function $\DZ_{mb}$ includes the $z$ boundaries of the \PYe
phase space, in particular those relevant to the global recoil employed
for the first final-state emission in MC@NLO matching. Equation~\eqref{eq:KMCnew}
is the direct implementation of Eq.~\eqref{eq:condkin}: the damping
interval is that of the scale which is actually written in the event
file, while the kinematic limit $\Qkin_{mb}$ is a separate veto. We
have verified numerically that the distribution of the effective dipole
scales obtained from Eqs.~\eqref{eq:Qab} and~\eqref{eq:Qdraw}
reproduces the product of the damping and of the kinematic cut that
enters Eq.~\eqref{eq:KMCnew}.

In the soft limit, the replacement of $\KMC$ by the soft FKS counterterm
through the ${\cal G}$ function~\cite{Frixione:2002ik,Alwall:2014hca} is
now multiplied by the average of the damping factors of the colour
connections of $m$, so that it also vanishes smoothly where the shower
cannot radiate. Previously, the replacement was not damped, and it could thus
extend beyond the phase space available to the shower.

\subsection{Initial-state emissions}
\label{sec:newISR}
The MC counterterms of an FKS sector are evaluated with the Born and
radiation variables of the parametrisation $\confH(\PhiB,\Phir)$ of the
real-emission phase space. In the parametrisation used so far in
\aNLOs~\cite{Frederix:2009yq}, the recoil of an initial-state emission
is shared by the two incoming partons, whose momentum fractions are both
rescaled. In \PYe, on the other hand, an initial-state emission changes
only the momentum fraction of the emitting parton, while the recoil is
absorbed by a Lorentz transformation of the hard final
state~\cite{Bierlich:2022pfr}. The two recoil strategies coincide only in the limit of soft or collinear radiation. We now adopt the same map as \PYe, see also Ref.~\cite{Torrielli:2010aw}: for an emission
off the incoming parton $j$, with $k$ the other incoming parton,
\beq
x_j=\frac{\bar{x}_j}{1-\xi}\,,\qquad x_k=\bar{x}_k\,,\qquad
0\le\xi\le1-\bar{x}_j\,,
\label{eq:ISRmap}
\eeq
where $\bar{x}_{j,k}$ and $x_{j,k}$ are the Born-level and real-emission
momentum fractions, and $\xi$ is defined in the real-emission partonic
c.m.~frame, so that $1-\xi$ is the momentum fraction $z$ of the \PYe
branching. The final-state momenta are obtained from the Born-level ones
with the Lorentz transformation employed by \PYe for initial-state
emissions off massless partons. Thus, the real-emission configuration
associated with given Born and radiation variables is the one that \PYe
generates when the corresponding branching is the first emission off the
$\clS$ event, and the MC counterterms are closer to what the shower does
away from the soft and collinear limits, in which the two maps coincide.
Since the phase-space measure and the soft and collinear limits are
unchanged, the analytically-integrated FKS counterterms only need to be
adapted to the new upper bound on $\xi$, and we have verified that
fixed-order cross sections are unchanged within the statistical
uncertainties. This map is now used by default, and it is the only one
available with shower matching.

\subsection{$\clH$-event starting scales}
\label{sec:newH}
The starting scale of $\clH$ events is
\beq
Q_\clH=\max\Big(Q^{(n+1)}_{ij},\,Q_{m\bar{b}}\Big)\,,
\label{eq:QH}
\eeq
where $Q_{m\bar{b}}$ is the effective scale, Eq.~\eqref{eq:Qab}, of the
dipole formed by the mother $m$ and one of its colour partners $\bar{b}$
in the sampled flow (chosen with equal probabilities if $m$ is a gluon),
computed with the same $Q$ as that of the $\clS$ contribution of the same
FKS sector and fold. The first argument is
\begin{eqnarray}
Q^{(n+1)}_{ij}&=&\max\Big(f f_2 R_{ij},\,f f_1 R_{ij}+\delta,\,\Qcut\Big)\,,
\nonumber
\\
R_{ij}&=&\min\big(M_{ij},\muref(\confH)\big)\,,
\label{eq:Qnpo}
\end{eqnarray}
with $M_{ij}$ the invariant mass of the FKS parton and its sister if they
are colour-connected in the $\clH$-event flow, and the smallest of the
invariant masses of $i$ or $j$ with their colour partners otherwise. In
the soft and collinear limits $R_{ij}\to 0$, and therefore $Q_\clH\to
Q_{m\bar{b}}$: soft and collinear $\clH$ events are showered in a similar
way to the $\clS$ events against which they are cancelled. For hard
emissions, $Q_\clH$ is determined by the hardness of the emission, but it
is never smaller than the corresponding $\clS$-event dipole scale.

\subsection{FxFx merging}
\label{sec:newFxFx}
In FxFx merging~\cite{Frederix:2012ps}, MC@NLO samples of different
multiplicities are combined, with the short-distance cross sections
reweighted by Sudakov factors in the CKKW
spirit~\cite{Catani:2001cc,Hamilton:2012np}, between the hard scale and
the smallest clustering scale $d_{\min}$, and the shower must take over
below the latter. In the new FxFx implementation, the starting scale of
$\clS$ events is therefore not randomised:
\beq
Q=f\,d_{\min}(\confS)\equiv f\muref(\confS)\,,
\label{eq:QFxFx}
\eeq
which coincides with the lowest scale of the Sudakov factors (for
$f=1$). Consistently, the MC counterterms are damped with
$D(q;Q,Q+\delta)$, i.e.~with a sharp damping function smoothed over the
narrow interval $\delta=3$~GeV for numerical reasons, and they are cut by
the \PYe kinematic limits as in Eq.~\eqref{eq:KMCnew}. The $\clH$-event
scales are given by Eq.~\eqref{eq:QH}, with $\muref(\confH)$ including
the clustering of the extra parton and with $Q_{m\bar b}=Q$.

The Sudakov factors of FxFx merging~\cite{Frederix:2012ps} are Sudakov
form factors~\cite{Dokshitzer:1978hw} at next-to-leading logarithmic
(NLL) accuracy~\cite{Catani:1991hj}, which include the two-loop
coefficient of the soft-gluon emission probability~\cite{Catani:1990rr}.
Their exponents are obtained by integrating numerically, over the scale
of the branching, the splitting kernels integrated over the momentum
fraction $z$. With \PYe matching, their exponents now include
finite terms: for massless partons, the $z$ integrals of the \PYe splitting
kernels are
performed exactly up to $z<1-\epsilon$, with
$\epsilon=\frac{r}{2}\big(\sqrt{4+r^2}-r\big)$ and $r=q/Q$. This bound
follows from the global-recoil phase space of \PYe for initial-state
emissions, $q$ being the scale of the branching and $Q$ the
upper scale of the Sudakov factor, identified with the dipole mass. This
retains the logarithmic coefficients, and the same finite terms enter the
${\cal O}(\as)$ expansion of the Sudakov factors that is subtracted from
the NLO cross sections, so that the NLO accuracy is unaffected. This is a
universal prescription, which does not include ratios of PDFs, the recoil
of the event or matrix-element corrections, and which is thus not the
no-emission probability of \PYe; it brings, however, the Sudakov
reweighting closer to the no-emission probability of the shower. It is
always used for FxFx merging with \PYe.

Finally, the FxFx clustering, which dominated the CPU time of the former
implementation, has been made considerably faster
(Section~\ref{sec:fxfxnegw}). The set of Feynman-diagram topologies
compatible with the clusterings performed so far is now tracked by means
of bit masks, so that checking whether a candidate clustering is allowed
requires a single bitwise operation, rather than a loop over all of the
diagrams and their propagators. Furthermore, the Sudakov factors, which
are integrated numerically and require many evaluations of $\as$, employ
an interpolation grid for the latter, and the clustering of the
real-emission configuration is computed once and reused for all of the
shower histories when the MC counterterms are summed over them
(Section~\ref{sec:sectors}).

\subsection{\aNLOD}
\label{sec:newDelta}
In \aNLOD~\cite{Frederix:2020trv}, the $\clH$-event short-distance cross
section of Eq.~\eqref{eq:xsecH} is multiplied by a factor $\Delta$, and
the complement, $(1-\Delta)(R-\KMC)$, is moved to the $\clS$ events. The
factor $\Delta$ is the product over the Born partons of their shower
no-emission probabilities, from the starting scales $\mu_{ab}$ of their
colour connections down to the stopping scales $t_{ab}(\confH)$,
i.e.~the values of the shower evolution variable associated with the
real-emission configuration. The $\clS$-event starting scales
thus enter $\Delta$ directly. In the new implementation, they are the
effective dipole scales of Eq.~\eqref{eq:Qab},
\beq
\mu_{ab}=Q_{ab}\,,
\label{eq:muab}
\eeq
obtained from the same single draw of $Q$ as in ordinary MC@NLO. These
scales are written in the event file as dipole-tagged starting
scales~\cite{Frederix:2020trv}, together with $\SCALUP=Q$. As a
consequence, the showering conditions of $\clS$ events are identical in
MC@NLO and \aNLOD, and the starting scales of the Sudakov factors in
$\Delta$ are precisely those which \PYe uses for the corresponding
dipoles. This replaces the independent dipole-by-dipole generation of
Ref.~\cite{Frederix:2020trv}, which also suffered from the $\Phir$
dependences discussed in Section~\ref{sec:old}. Likewise, the colour flow
used in the construction of $\Delta$ is the one sampled at the Born level
(Section~\ref{sec:Bornlevel}), rather than one chosen according to rule
{\em R.2} of Ref.~\cite{Frederix:2020trv}. A connection is treated as a
dead zone of $\Delta$ also when its stopping scale exceeds its starting
scale. The $\clH$-event starting scales are set equal to the stopping
scales, as in Ref.~\cite{Frederix:2020trv}.

The stopping scales and dead zones are now computed by a Fortran
transcription of the corresponding \PYe routines (version~8.313), which
belong to its CKKW-L merging implementation~\cite{Lonnblad:2011xx}. The
hard-process executable of \aNLOs thus no longer needs to be linked with
\PYe at run time; \PYe is still used to pre-tabulate the Sudakov
factors~\cite{Frederix:2020trv}. This considerably reduces the cost of
\aNLOD: while in Ref.~\cite{Frederix:2020trv} it increased the running
time by a factor of 2 to 3 w.r.t.~ordinary MC@NLO, the increase is now at
most 20\% for the processes of Table~\ref{tab:options}. This transcription and the one of the
starting scales of Eq.~\eqref{eq:Qab} have been validated against \PYe:
$5.6\cdot10^4$ evaluations of individual scales agree exactly, and
$3.5\cdot10^3$ complete evaluations for seven processes, which include
massive emitters and double colour connections, agree to better than
$10^{-10}$ (in units of $\max(1~{\rm GeV},|{\rm scale}|)$).

\subsection{\HWs}
\label{sec:newHW}
Although we focus on \PYe in this paper, we mention that the same
construction is employed for matching with the angular-ordered shower of
\HWs~\cite{Bellm:2015jjp,Bewick:2023tfi}, where \SCALUP acts as a veto on
the transverse momentum of the emissions, independently of the
angular-ordering starting scales. A single damped scale $Q$ is written in
the event file, and the effective dipole scales are
$\min(Q,\pT^{\max})$, with $\pT^{\max}$ the largest transverse momentum
allowed by the angular starting scales of each colour
connection~\cite{Gieseke:2003rz} (e.g.~one quarter of the maximal
angular variable for a massless final-state emitter). These have been
validated against the corresponding \HWs (version~7.3.0) routines, with
which they agree to better than $10^{-12}$ (in the same units as above). \aNLOD remains available
for \PYe only.

\section{New possibilities}
\label{sec:opp}
\begin{table}[t]
\begin{center}
\small
\begin{tabular}{lll}
\toprule
label & input parameter & Section \\
\midrule
\vnone   & ---                               & ---                          \\
\vsumkl  & \param{MCExplicitKLSum}           & \ref{sec:sectors}           \\
\vsumij  & \param{FKSExplicitSum}            & \ref{sec:sectors}           \\
\vmcflow & \param{MCSubtractionAtFixedFlow}  & \ref{sec:sectors}           \\
\vbs     & \param{born\_spreading}          & \ref{sec:BS}                \\
\vdelta  & \param{mcatnlo\_delta}           & \ref{sec:newDelta}          \\
\vall    & all of the above                  &                              \\
\bottomrule
\end{tabular}
\caption{Labels of the options of the new implementation used in
Sections~\ref{sec:res} and~\ref{sec:resFxFx}, with the input parameters
that switch them on (in \param{FKS\_params.dat} for the first three
options, and in the run card for the last two) and the sections where
they are described. \vnone and \vall denote the new implementation with
none and all of the options switched on, respectively.}
\label{tab:labels}
\end{center}
\end{table}
\begin{table*}[t]
\begin{center}
\begin{tabular}{lrrrrrrr}
\toprule
process & \vnone & \vsumkl & \vsumij & \vmcflow & \vbs & \vdelta & \vall \\
\midrule
& \multicolumn{7}{c}{fraction $f$ of negative-weight events} \\
\cmidrule(lr){2-8}
$pp\to e^+e^-$ & 6.2\% & 5.3\% & 5.0\% & 6.1\% & 3.3\% & 4.6\% & 0.8\% \\
$pp\to H$ & 7.0\% & 5.4\% & 6.9\% & 7.1\% & 3.9\% & 4.4\% & 0.2\% \\
$pp\to t\bt$ & 22.6\% & 20.6\% & 21.2\% & 22.6\% & 19.7\% & 14.4\% & 8.0\% \\
$pp\to t\bt W^+$ & 15.0\% & 14.0\% & 14.1\% & 15.0\% & 14.2\% & 10.2\% & 8.1\% \\
\midrule
& \multicolumn{7}{c}{CPU time} \\
\cmidrule(lr){2-8}
$pp\to e^+e^-$ & 8m40s & 12m59s & 11m22s & 9m06s & 20m02s & 9m38s & 52m54s \\
$pp\to H$ & 14m59s & 22m04s & 23m04s & 14m34s & 18m07s & 16m36s & 49m46s \\
$pp\to t\bt$ & 49m43s & 1h54m & 1h42m & 38m00s & 1h05m & 59m48s & 4h10m \\
$pp\to t\bt W^+$ & 1h54m & 4h05m & 2h12m & 1h58m & 2h09m & 1h47m & 10h19m \\
\bottomrule
\end{tabular}
\caption{Fraction $f$ of negative-weight events (upper half) and total
CPU time, summed over all jobs (lower half), for samples of $10^6$
events at the 13~TeV LHC, generated in the five-flavour scheme with the
default run card, the \PYe MC counterterms and no cuts on jets. Columns:
with none of the options (\vnone); with the explicit sum over histories
of Eq.~\eqref{eq:Hkl} (\vsumkl); with the explicit sum over FKS sectors
before the unweighting of $\clS$ events (\vsumij); with the MC
counterterms evaluated at the sampled colour flow (\vmcflow); with Born
spreading (\vbs); with \aNLOD (\vdelta); with all of them (\vall); see
also Table~\ref{tab:labels}.}
\label{tab:options}
\end{center}
\end{table*}

\subsection{MC counterterms, FKS sectors and colour flows}
\label{sec:sectors}
In the former implementation, the starting scale of an $\clS$ event was
determined during the evaluation of the MC counterterms of a given FKS
sector, and depended on the real-emission configuration sampled in that
sector and on the colour partners of its FKS pair. Therefore, the
starting scales were tied to the way in which the MC counterterms are
organised in sectors. With the new construction, the starting scales
and the colour flow are attributes of the Born configuration alone, and
the MC counterterm of any shower history can be evaluated consistently
for any Born configuration. This has allowed us to investigate the
following modified formulations of the matching, each of which can be
switched on independently. Table~\ref{tab:labels} lists the corresponding
input parameters, together with the short labels by which we refer to
these options, as well as to Born spreading and \aNLOD, in
Sections~\ref{sec:res} and~\ref{sec:resFxFx}.

\paragraph{Explicit sum over shower histories (\vsumkl).}
In \aNLOs, the phase space of the real emission is partitioned by means
of the FKS $\Sfun_{ij}$ functions, with \mbox{$\sum_{ij}\Sfun_{ij}=1$}, and
the $\clH$-event integrand of sector $ij$ is
\mbox{$\Sfun_{ij}R-\KMC_{ij}$}, where $\KMC_{ij}$ describes only the shower
branchings in which $i$ is emitted off $j$, and is not multiplied by
$\Sfun_{ij}$~\cite{Alwall:2014hca}. The real emission and the
counterterms are thus damped differently away from the singular limits
of the sector, which induces negative weights, e.g.~in the anticollinear
configurations discussed in Ref.~\cite{Frederix:2020trv}. In the new
formulation, the first shower emission is decomposed into its histories
$kl$, and the complete counterterm is partitioned in the same way as the
real emission, as anticipated in Ref.~\cite{Alwall:2014hca}:
\beq
d\sigma^{(\clH)}_{ij}=\Sfun_{ij}\sum_{kl}\Delta_{kl}
\Big(\Sfun_{kl}R-\KMC_{kl}\Big)\,d\PhiB\,d\Phir\,,
\label{eq:Hkl}
\eeq
with $\Delta_{kl}=1$ in MC@NLO, and $\Delta_{kl}$ the no-emission
probability of history $kl$ in \aNLOD. Each history is evaluated with its
own Born configuration $\PhiB^{(kl)}$ and radiation variables
$\Phir^{(kl)}$, obtained by inverting the kinematic map of its sector.
Its damping interval, Eq.~\eqref{eq:mu12} with
$\muref(\confS(\PhiB^{(kl)}))$, its dead zones and $\Delta_{kl}$ are
computed from these variables as well. The result is then converted into
the integration measure of sector $ij$, which determines the kinematics,
colour flow and starting scale of the $\clH$ event. The $\clS$ events retain the counterterm of each history in its
own variables, which gives the same integral. Since
\mbox{$\sum_{ij}d\sigma^{(\clH)}_{ij}$} is unchanged, so is the unshowered
cross section; however, in MC@NLO the $\clH$-event weights are now
negative essentially only where the complete MC counterterm exceeds the
real-emission matrix element. Clearly, Eq.~\eqref{eq:Hkl} can be consistently constructed only
if the damping of $\KMC_{kl}$ is a known function of $\PhiB^{(kl)}$,
which is precisely what Eqs.~\eqref{eq:solution}--\eqref{eq:mu12}
guarantee. This formulation is also better suited for showers whose
branchings do not map onto single FKS pairs.

\paragraph{Monte Carlo over colour flows (\vmcflow).}
As discussed in Section~\ref{sec:Bornlevel}, the MC counterterms can be
evaluated either at the sampled colour flow only, or by summing
explicitly over the leading colour flows, i.e.~by using
Eq.~\eqref{eq:KMC0} with the damping and kinematic limits of
Eq.~\eqref{eq:KMCnew} for each flow. In both cases the event inherits the
sampled flow, and the two options, the former of which we denote by
\vmcflow, differ only in the variance of the integrand.

\paragraph{Explicit sum over FKS sectors (\vsumij).}
Rather than sampling one (group of) FKS sector(s) per integration point,
all sectors can be summed explicitly at each Born configuration, before
the absolute value of the $\clS$-event weights is taken for the
unweighting. In this way, cancellations between the $\clS$-event
contributions of different sectors take place before the sign of the
weight is determined, which complements folding and Born spreading, at
the price of more matrix-element evaluations per point. A common Born
phase-space parametrisation is used, so that all of the sectors share
the same Born configuration. Combining the contributions of different
sectors into a single $\clS$ event is meaningful only if they share the
same colour flow and starting-scale distribution, which was not the case
with the former implementation.

\subsection{Born spreading}
\label{sec:BS}
Born spreading~\cite{Frederix:2023hom} exploits the fact that the Born
contribution to Eq.~\eqref{eq:xsecS} does not depend on $\Phir$, and can
thus be redistributed over the radiation phase space with any normalised
function, $\rho(\Phir|\PhiB)$ in the notation of Section~\ref{sec:cond}:
\beq
d\sigmaS=\Big[B\rho+\frac{V}{\int d\Phir}+\KMC-C\Big]\,d\PhiB\,d\Phir\,,
\label{eq:BS}
\eeq
with $\int d\Phir\,\rho=1$.
The function $\rho$ is chosen so as to concentrate the Born contribution
where the rest of the integrand is negative, thereby reducing the
fraction of negative-weight $\clS$ events at little computational
cost. Since $\clS$ events have Born kinematics, their distributions are
unaffected, provided that nothing that is passed to the shower depends
on $\Phir$~\cite{Frederix:2023hom}. As shown in Section~\ref{sec:cond},
this is the very condition for the consistency of MC@NLO. Therefore,
with the new construction Born spreading introduces no bias, and it has
been included in the public version of the code, where it is switched on
by means of the \param{born\_spreading} parameter of the run card (\vbs
in Table~\ref{tab:labels}).

In the public implementation, $\rho$ is a table of $40\times40$ bins in
the radiation variables $u$ and $v$ in which the Born measure is flat
(i.e.~the squares of the integration variables associated with $\xi$ and
$y$, the former rescaled by its upper bound in the case of massive
final-state emitters), independent of $\varphi$. A separate table is used
for each FKS sector, and each table is normalised separately, which
preserves the Born contribution at fixed $\PhiB$. In the grid-setup
stage, after a first adaptation of the integration grids, the tables are
determined by sampling the integrand at $8\cdot10^5$ points and by
minimising, under the normalisation constraint, the sum of the negative
parts of the sampled $\clS$-event weights. For each sector this is a separable, convex,
piecewise-linear minimisation problem, which is solved exactly. The
residual part of the virtual corrections, which in \aNLOs is evaluated
separately from the Born-based approximation of the
virtual~\cite{Alwall:2014hca}, does not depend on $\rho$, and is not
included in the fit. The tables are retained only if they reduce the
negative-weight sum by more than two standard deviations on $2\cdot10^5$
independent points; the integration grids are then re-adapted to the
spread integrand. The overhead is thus a fixed cost, independent of the
number of events, whereas folding multiplies the evaluation of the real
emission and of the counterterms for every event by the number of folds.
This construction supersedes the baseline spreading function of
Ref.~\cite{Frederix:2023hom}, proportional to the negative part of the
non-Born $\clS$ integrand averaged over $\PhiB$, and the improved
non-iterative and iterative variants studied in Ref.~\cite{Che:2024bsc},
in that it directly minimises the quantity of interest. Its effect on the
fraction of negative weights is discussed in Section~\ref{sec:negw}.

Since in FxFx merging each multiplicity is an MC@NLO sample with
Born-level starting scales, Eq.~\eqref{eq:QFxFx}, Born spreading can now
be used also in conjunction with FxFx merging, which was not possible
previously, owing to the dependence of the FxFx $\clS$-event scales on the
real-emission configuration. This combination is studied in
Section~\ref{sec:resFxFx}.

\section{Results}
\label{sec:res}
In this section, we present results obtained with the new construction
for four processes at the 13~TeV LHC, matched to \PYe: Drell-Yan
($pp\to e^+e^-$), Higgs ($pp\to H$), $t\bt$ and $t\bt W^+$ production.
For each of them, samples of $10^6$ events are generated with the variants
of Table~\ref{tab:options} and with the former implementation
(\aNLOs version~3.5.18, the current long-term-stable release), in the five-flavour scheme, with the default run card of \aNLOs, the \PYe MC
counterterms, no cuts on jets, and the NNPDF4.0 NLO PDFs with
$\as(m_Z)=0.118$~\cite{NNPDF:2021njg}, accessed through
LHAPDF~\cite{Buckley:2014ana}. The events are
showered with \PYe version~8.317, with pure QCD showers: multi-parton
interactions, hadronisation, QED radiation and primordial transverse
momentum are switched off, and the top quarks and $W$ bosons are kept
stable.

\subsection{Negative weights and CPU time}
\label{sec:negw}
Table~\ref{tab:options} reports the fraction $f$ of negative-weight events
and the CPU time with each of the options of Section~\ref{sec:sectors}
switched on separately, with Born spreading (Section~\ref{sec:BS}), with
\aNLOD (Section~\ref{sec:newDelta}), and with all of them combined. We recall that the number of events needed for
a given statistical precision is proportional to
$(1-2f)^{-2}$~\cite{Frederix:2020trv}. On their own, the sum over
histories and the sum over FKS sectors lower $f$ by up to two percentage
points, for 1.2 to 2.3 times the CPU time, while evaluating the MC
counterterms at the sampled colour flow leaves $f$ unchanged. The latter
option matters only for $t\bt$ production, the only one of these
processes with more than one leading colour flow at the Born level, for
which it lowers the CPU time by about a quarter. Born spreading nearly
halves $f$ for Drell-Yan and Higgs production, and lowers it by up to
three percentage points for the processes with top quarks, for which a
large part of the negative weights is associated with $\clH$ events. Its
extra CPU time, between 13\% and 31\% except for Drell-Yan production, is
mainly due to the fit of $\rho$ and to the second adaptation of the
integration grids. This is a fixed cost, which is large compared with the
short Drell-Yan run, and whose share decreases for larger samples. \aNLOD lowers $f$ by 1.5 to 5 percentage points, except for $t\bt$
production, where it brings it from 22.6\% down to 14.4\%. When all of the options are
combined with each other, and with Born spreading and \aNLOD, $f$ is reduced more than by
any of them alone: below 1\% for Drell-Yan and Higgs production, and to 8.0\% and 8.1\%
for $t\bt$ and $t\bt W^+$ production, respectively, for 3 to 6 times the
CPU time. For $t\bt$ production, a given statistical precision then
requires 2.4 times fewer events, which more than compensates the
longer generation once the shower and the detector simulation of the
events are included.

\subsection{Transverse momentum of the hard system}
\label{sec:hardpt}
\begin{figure*}[tp]
\begin{center}
\includegraphics[width=0.48\textwidth]{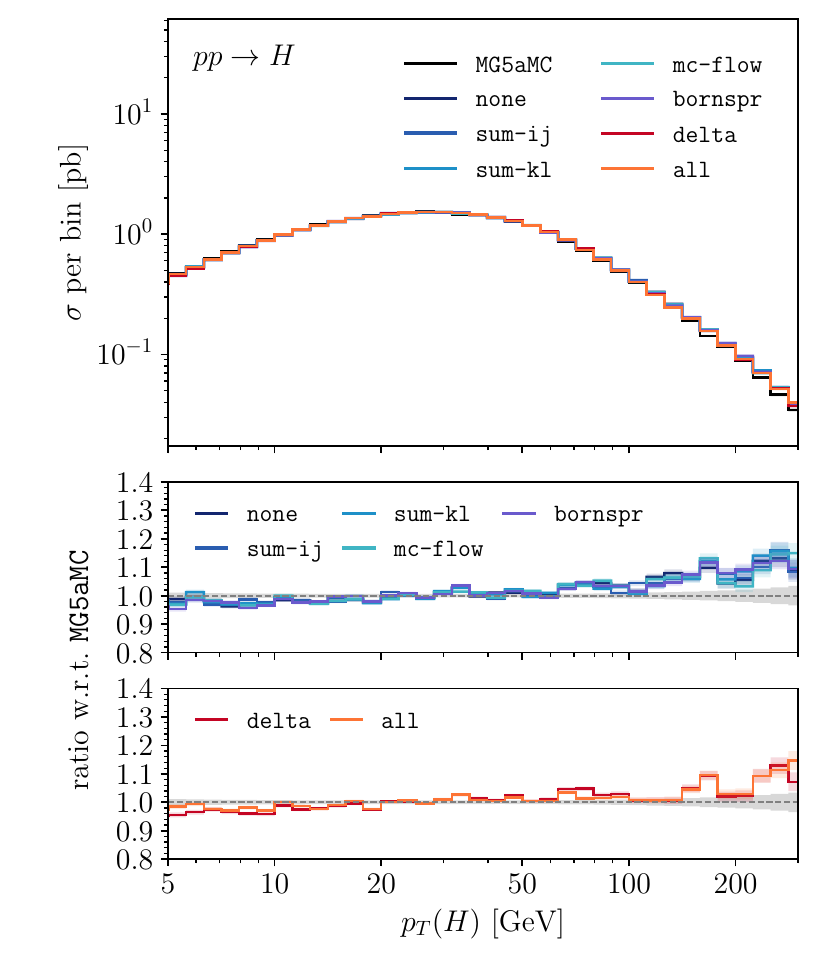}
\hfill
\includegraphics[width=0.48\textwidth]{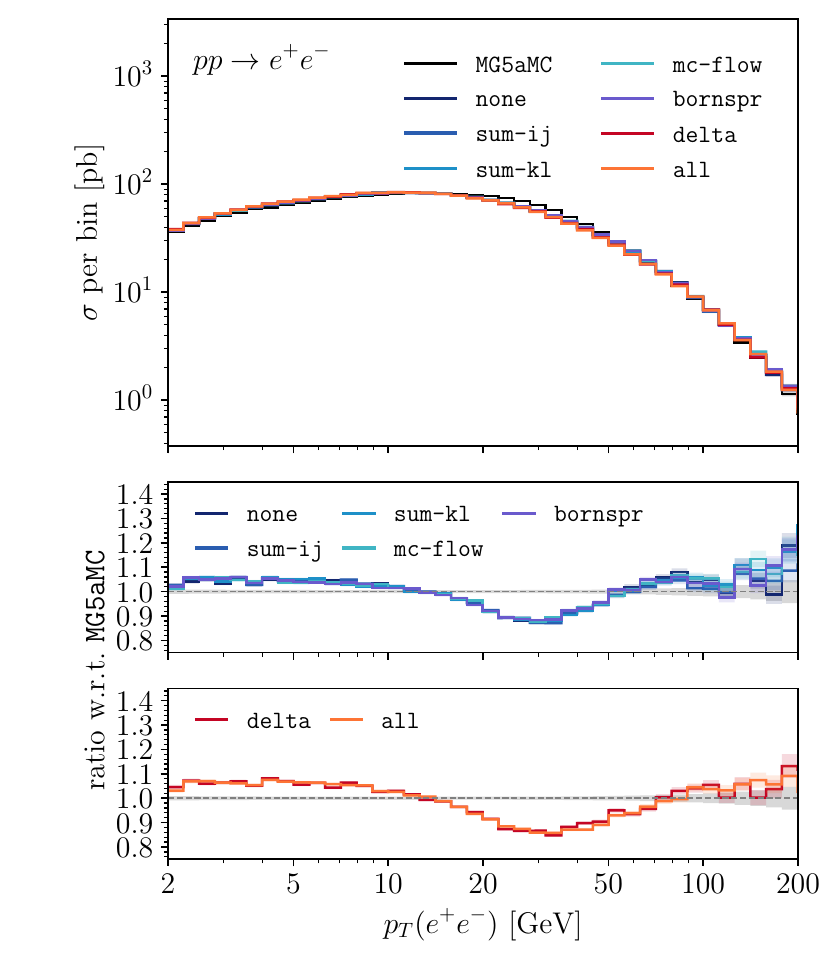}

\medskip
\includegraphics[width=0.48\textwidth]{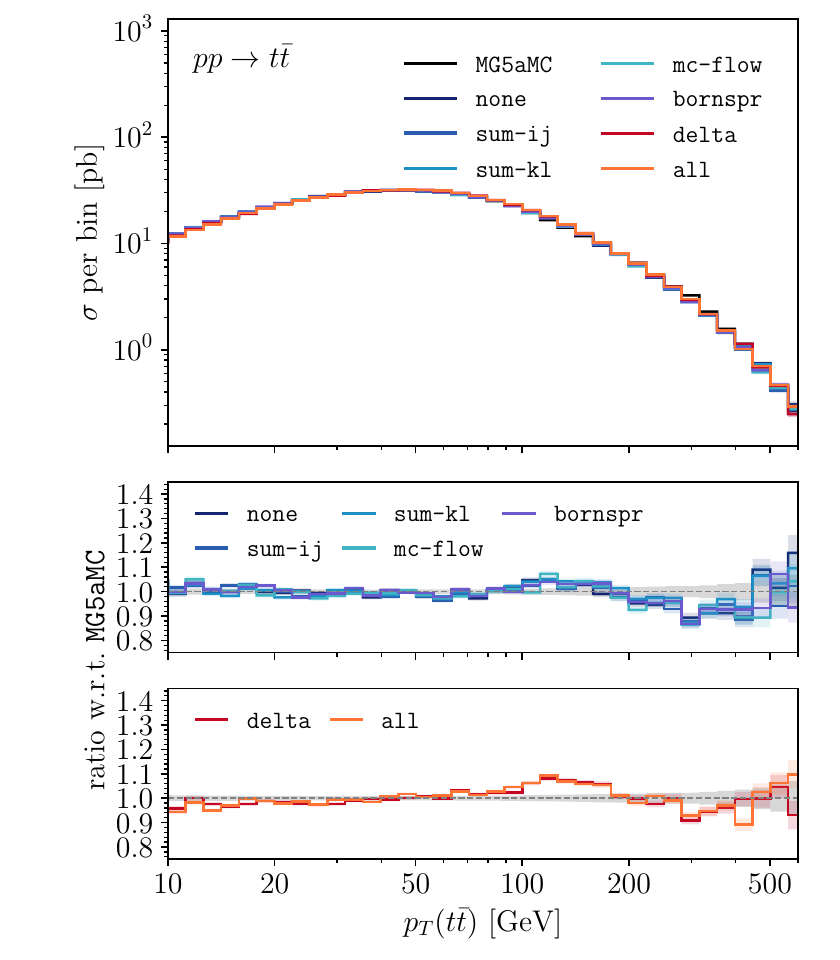}
\hfill
\includegraphics[width=0.48\textwidth]{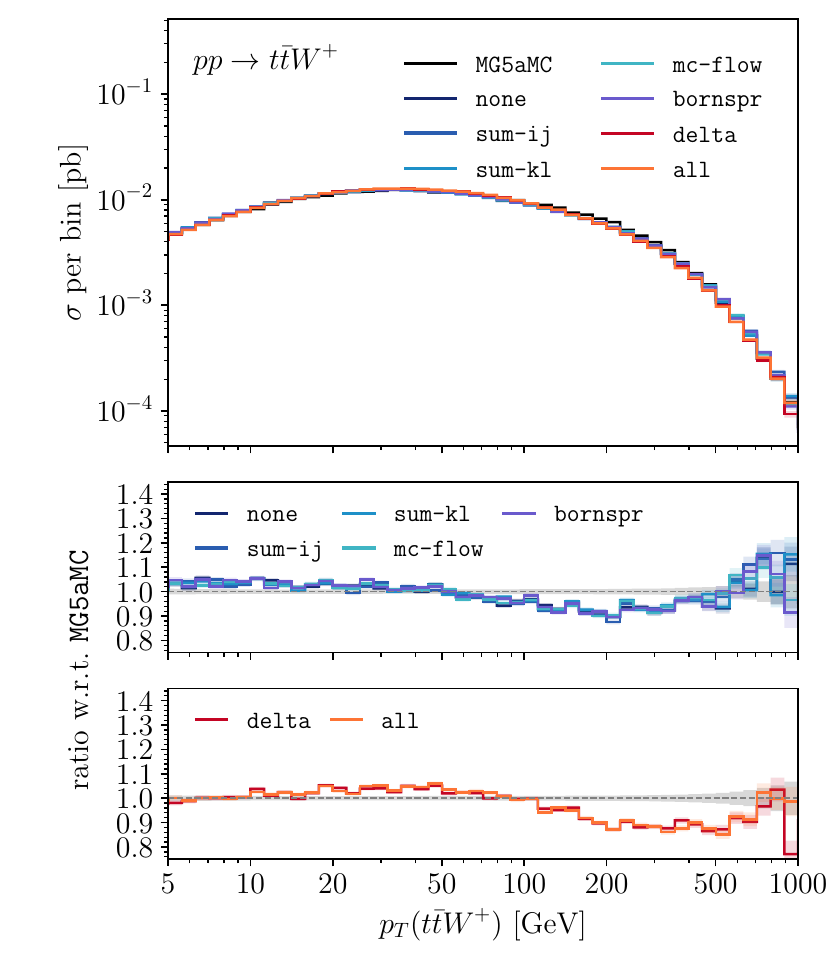}
\caption{Transverse momentum of the hard system in Higgs
(top left), Drell-Yan (top right), $t\bt$ (bottom left) and $t\bt W^+$
(bottom right) production at the 13~TeV LHC, after the \PYe shower, for
the variants of Table~\ref{tab:options} and for the former implementation
(\texttt{MG5aMC}). In each plot, the upper panel shows the cross section
per bin, with bins uniform in $\log_{10}(\pT/\mathrm{GeV})$, while the
middle and lower panels show the ratios of the variants without and with
\aNLOD, respectively, to the former implementation (without \aNLOD). The
bands show the statistical uncertainties.}
\label{fig:hardpt}
\end{center}
\end{figure*}

In Fig.~\ref{fig:hardpt} we show the distributions of the transverse
momentum of the hard system in Higgs, Drell-Yan, $t\bt$
and $t\bt W^+$ production, after the \PYe shower. These transverse
momenta vanish at the Born level, and the distributions are therefore
sensitive to the shower and to its matching to the fixed-order
calculation. The options other than \aNLOD do not change the predictions
at NLO, and the corresponding variants (middle panels) agree with each
other within the statistical uncertainties. We do not show theoretical
uncertainties, e.g.~from scale variations. The differences of these
variants w.r.t.~the former implementation stem from the new
construction as a whole, including the map for initial-state emissions of
Section~\ref{sec:newISR}. These differences are
process dependent: they amount to a few percent over most of the Higgs
and $t\bt$ distributions, and reach 10--12\% in well-populated regions of
the Drell-Yan and $t\bt W^+$ distributions. In particular, the Drell-Yan
distribution is lowered by about 12\% around
$\pT(e^+e^-)\simeq30$~GeV. This difference arises mainly from the 30~GeV
lower bound on $\mu_2$ of the former implementation (items {\em a)} and
{\em b)} in Section~\ref{sec:old}), which is absent from the new
construction. In the high-$\pT$ tails, the new predictions lie above the
former ones by up to about 10\% for Higgs and Drell-Yan production. This
is primarily due to the larger starting scales assigned to hard $\clH$
events in the new construction, Eq.~\eqref{eq:QH}, which were bounded by
$\mu_2$ in the former implementation. For $t\bt$ and $t\bt W^+$
production, the new predictions are instead lower than the former ones by
5--10\% between 200 and 450~GeV and between 110 and 500~GeV,
respectively, and they rise above them only at the largest transverse
momenta, consistently with the larger values of $\mu_2$ for these
processes: by about 9\% above 700~GeV for $t\bt W^+$ production, and
possibly also in the last bins of the $t\bt$ distribution, where the
statistical uncertainties are however large.

\aNLOD (lower panels) changes the predictions only beyond NLO. Since the
ratios in the lower panels are also taken w.r.t.~the former
implementation without \aNLOD, they include the effects of the new
construction as well as those of \aNLOD. The latter are given by the
ratio of the \vdelta and \vnone predictions, both of which are obtained with
the new construction: \aNLOD lowers the Higgs distribution by up to 7\%
between 110 and 140~GeV, and the Drell-Yan one by up to 8\% between 50
and 90~GeV; it raises the $t\bt$ and $t\bt W^+$ distributions by up to 7\%,
between 50 and 280~GeV and between 20 and 110~GeV, respectively, and
lowers the $t\bt W^+$ one by 5--10\% above 250~GeV. The results obtained
with all options combined agree with those of \aNLOD alone within the
statistical uncertainties.

The differences between ordinary MC@NLO and \aNLOD are thus of similar
size for the four processes. Compared with Figs.~2--5 of
Ref.~\cite{Frederix:2020trv}, they are much smaller for $t\bt$
production, where the enhancement of \aNLOD at intermediate $\pT$
reached about 30\% in that reference, whereas they are somewhat larger
for Higgs and Drell-Yan production, for which the two predictions were
found to be within $\pm5\%$ of each other there. For $t\bt W^+$
production, the \aNLOD distribution is now softer than the ordinary
MC@NLO one at large $\pT$, whereas a harder tail was found in
Ref.~\cite{Frederix:2020trv}. Among other things, the two studies differ
in their scale prescriptions: in the present construction, the $\clS$
events are showered under the same conditions in MC@NLO and \aNLOD, as
discussed in Section~\ref{sec:newDelta}.

\section{\texorpdfstring{$W$}{W}+jets production with FxFx merging}
\label{sec:resFxFx}
FxFx merging is where the problems of the former implementation were
most acute: the $\clS$-event starting scale was often smaller than the
lowest scale of the Sudakov reweighting (Section~\ref{sec:old}), and Born
spreading could not be used at all.

\subsection{Setup}
\label{sec:fxfxsetup}
We study $pp\to e^+\nu_e$ production with up to two jets at NLO, which is
representative of the merged samples used at the LHC. We have generated
FxFx-merged samples with the three multiplicities
$pp\to e^+\nu_e+0,1,2$ partons at the 13~TeV LHC, with the NNPDF4.0 NLO
PDFs~\cite{NNPDF:2021njg} with $\as(m_Z)=0.118$ from LHAPDF~\cite{Buckley:2014ana}, the default dynamical scale and a generation
cut $\pT(j)>10$~GeV on the jets, reconstructed with the $k_T$
algorithm~\cite{Catani:1993hr} with $R=1$. For each variant,
$10^7$ events are generated on a single machine with 64 cores and showered with \PYe
version~8.317, again with pure QCD showers, without multi-parton
interactions, hadronisation, QED radiation and primordial transverse
momentum; the FxFx
veto~\cite{Frederix:2012ps} is then applied with merging scales
$\mu_Q=20$, 30 and 40~GeV. We compare the former implementation (\aNLOs version~3.5.18, the current long-term-stable release) with the
new one, without any of the options of Section~\ref{sec:sectors} (\vnone),
and with each of them, or Born spreading, switched on separately, as well
as with all of them combined (\vall); since \aNLOD is not yet available
with FxFx merging, \vall does not include it here.

\subsection{Merging efficiency, negative weights and CPU time}
\label{sec:fxfxnegw}

\begin{table*}[t]
\begin{center}
\begin{tabular}{lrrrrrrrrr}
\toprule
& & & & \multicolumn{2}{c}{$\mu_Q=20$~GeV} & \multicolumn{2}{c}{$\mu_Q=30$~GeV} & \multicolumn{2}{c}{$\mu_Q=40$~GeV} \\
\cmidrule(lr){5-6}\cmidrule(lr){7-8}\cmidrule(lr){9-10}
variant & CPU time & wall time & $f_{\rm LHE}$ & $\epsilon$ & $f$ & $\epsilon$ & $f$ & $\epsilon$ & $f$ \\
\midrule
\vnone    & 190~h & 3.9~h & 25.0\% & 49.3\% & 18.6\% & 44.6\% & 15.8\% & 42.4\% & 14.3\% \\
\vsumkl  & 437~h & 9.3~h & 22.8\% & 50.2\% & 16.7\% & 46.1\% & 14.1\% & 44.1\% & 12.8\% \\
\vsumij  & 304~h & 7.1~h & 22.6\% & 50.3\% & 16.2\% & 45.9\% & 13.5\% & 43.8\% & 12.1\% \\
\vmcflow & 155~h & 3.2~h & 24.9\% & 49.4\% & 18.6\% & 44.6\% & 15.7\% & 42.4\% & 14.2\% \\
\vbs     & 213~h & 4.2~h & 22.9\% & 49.0\% & 15.7\% & 44.3\% & 12.5\% & 42.1\% & 10.9\% \\
\vall    & 774~h & 16.1~h & 17.7\% & 52.0\% & 11.5\% & 48.9\% & 9.5\% & 47.4\% & 8.6\% \\
\midrule
\texttt{MG5aMC} & 1133~h & 22.1~h & 27.8\% & 47.3\% & 21.0\% & 40.6\% & 16.4\% & 37.3\% & 13.7\% \\
\bottomrule
\end{tabular}
\caption{FxFx-merged $pp\to e^+\nu_e+0,1,2$ jets at NLO, with a generation
cut $\pT(j)>10$~GeV, for samples of $10^7$ events: total CPU time summed over all jobs, and wall time on a
single machine with 64 cores (both without compilation and showering); fraction of
negative-weight events in the event file, $f_{\rm LHE}$; merging
efficiency $\epsilon$, i.e.~fraction of the events that pass the FxFx
veto after the \PYe shower, and fraction of negative-weight events among
them, $f$, for three merging scales $\mu_Q$. The variants are labelled as
in Table~\ref{tab:labels}, with \vall not including \aNLOD; \texttt{MG5aMC} is the
former implementation.}
\label{tab:fxfx}
\end{center}
\end{table*}

The results are collected in Table~\ref{tab:fxfx}. The merging
efficiency decreases\footnote{Since the fraction $f$ of negative weights after merging decreases
with increasing $\mu_Q$, and all events have the same absolute weight,
fewer accepted events are needed to reproduce the same cross section.} from about 50\% for $\mu_Q=20$~GeV to 42--44\% for
$\mu_Q=40$~GeV, and it is practically the same for the variants of the new
implementation with a single option. With all of them combined, it is
larger, 47--52\%, whereas with the former implementation it is lower than
for \vnone, by 2 to 5 percentage points, the difference growing with
$\mu_Q$. After merging, the fraction of negative weights is
considerably smaller than in the event file, and it decreases with
$\mu_Q$: the vetoed events stem mainly from the one- and two-parton
samples, whose negative-weight fractions after merging, about 25\% and
35\% without any option, are much larger than that of the zero-parton
sample, about 10\%. Born spreading is the most effective option, lowering
$f$ by 3 to 3.5 percentage points for all merging scales, mostly in the
zero-parton sample, where the negative weights are dominated by
$\clS$ events (for $\mu_Q=20$~GeV, from 10.0\% to 5.8\%). The sum over
FKS sectors and the sum over histories lower $f$ by 2 to 2.5 and by 1.5
to 2 percentage points, respectively, while evaluating the MC
counterterms at the sampled colour flow leaves $f$ unchanged, but reduces
the CPU time by about 20\%. With all of the options combined, $f$ drops to
11.5\%, 9.5\% and 8.6\% for $\mu_Q=20$, 30 and 40~GeV, respectively,
i.e.~by 6 to 7 percentage points w.r.t.~\vnone.

The gain of Born spreading, and of any other method that acts on the
$\clS$ events only, is limited by the $\clH$ events. Without any option,
these are only 16--17\% of the merged events, but about half of them have
a negative weight, so that they account for 51--53\% of the negative
weights after merging, while the fraction of negative-weight $\clS$
events is 8--11\%. Born spreading lowers the latter to 3--6\%, leaving
the former unchanged, so that 67--76\% of the remaining negative weights
stem from $\clH$ events. Furthermore, Born spreading is much less
effective in the two-parton sample: for $\mu_Q=30$~GeV, it lowers the
fraction of negative-weight $\clS$ events from 6.8\% to 1.8\% in the
zero-parton sample and from 13.0\% to 8.1\% in the one-parton sample,
but only from 20.0\% to 17.7\% in the two-parton one. There, the sum over
FKS sectors is more effective, with 14.9\%, and all of the options
combined bring this fraction down to 10.5\%. With all of the options, the
fraction of negative-weight $\clS$ events after merging is 2--4\%, and
70--77\% of the remaining negative weights stem from $\clH$ events. A
further reduction requires addressing the $\clH$ events, e.g.~with \aNLOD. We discuss this perspective in Section~\ref{sec:conc}.

Taking into account both the merging efficiency and the negative
weights, the number of events in the event file needed for a given
statistical precision after merging is proportional to
$[\epsilon(1-2f)^2]^{-1}$. For $\mu_Q=30$~GeV, the CPU time per $10^6$
effective merged events is then 91~h without options, 85~h with Born
spreading, 74~h with the sampled colour flow, 124~h with the sum over FKS
sectors, 184~h with the sum over histories, and 241~h with all of them
combined. In the latter case, each merged event, i.e.~each event that is
passed to the detector simulation, carries 1.4 times the statistical
power of an event of \vnone, which can compensate the higher generation
cost once that of the detector simulation is included. With the former
implementation, which took about six times longer to generate the
samples, in terms of both CPU and wall time, it is 617~h. With the former
implementation, 83\% of the CPU time is spent in the FxFx clustering and
in the computation of the Sudakov factors, i.e.~about 950~h, against about
70~h with the new one: the speed-up is thus mostly due to the faster
clustering described in Section~\ref{sec:newFxFx}.

\subsection{Distributions}
\label{sec:fxfxdist}

Born spreading, the sum over FKS sectors and the sampled colour flow leave
the merged distributions unchanged: over the 21 inclusive observables of
our analysis (transverse momenta and rapidities of the $e^+\nu_e$ pair,
of the positron and of the three hardest jets, jet separations and the
differential jet rates $d_{01}$ to $d_{34}$), their $\chi^2$ per bin
w.r.t.~\vnone is between 0.99 and 1.13 for all merging scales.
In particular, this confirms that Born spreading can be used in
conjunction with FxFx merging. The sum over histories also agrees with
\vnone, except for the $d_{23}$ and $d_{34}$ jet rates, most markedly for
$\mu_Q=20$~GeV; we attribute this beyond-NLO effect to the different
assignment of the $\clH$ events to the FKS sectors, and thus to their
starting scales. The predictions with all of the options combined agree
with those of \vsumkl within the statistical uncertainties ($\chi^2$ per
bin between 1.02 and 1.08), and thus share these differences.

\begin{figure*}[tp]
\begin{center}
\includegraphics[width=0.48\textwidth]{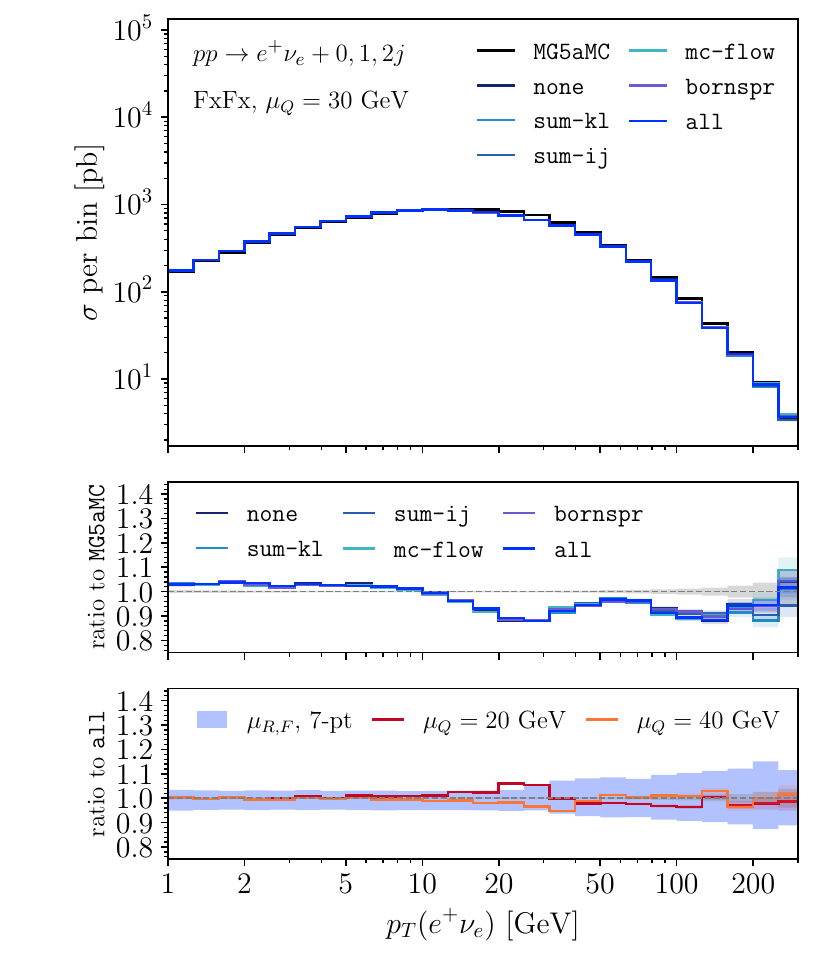}
\hfill
\includegraphics[width=0.48\textwidth]{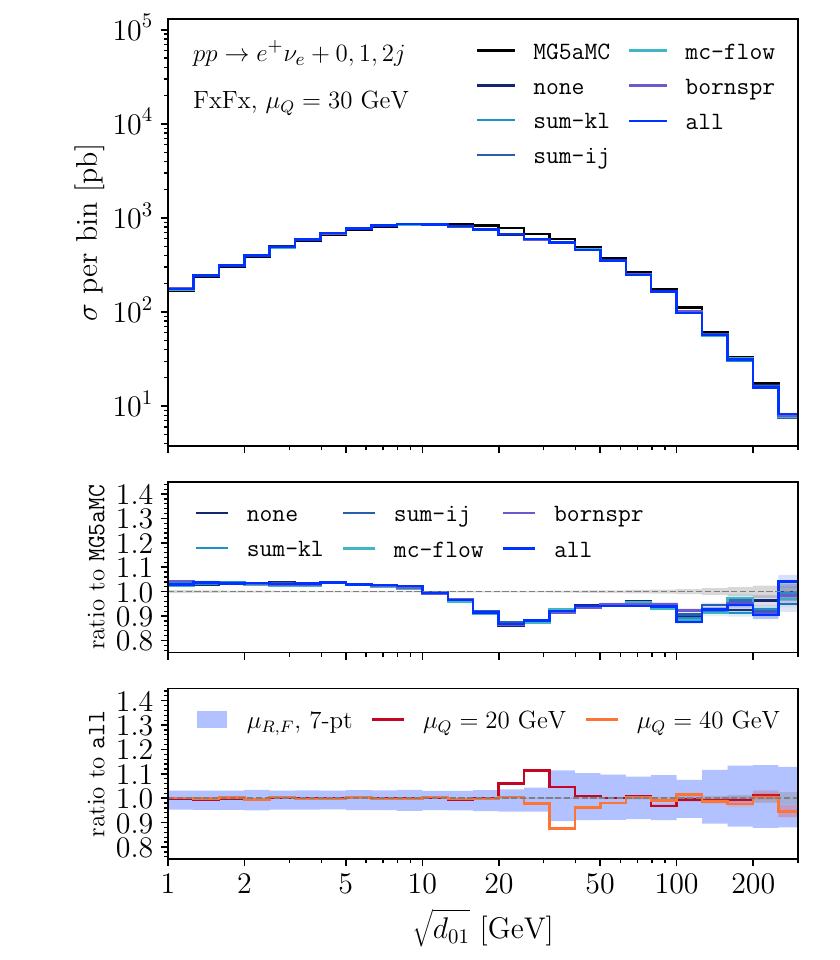}

\medskip
\includegraphics[width=0.48\textwidth]{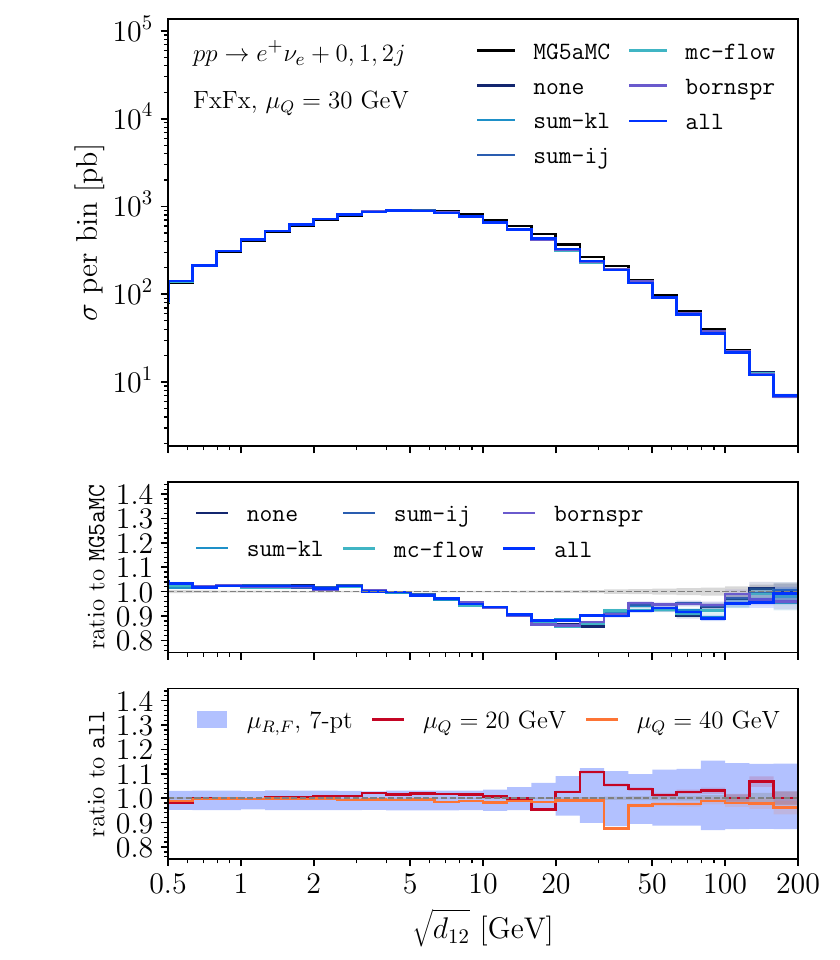}
\hfill
\includegraphics[width=0.48\textwidth]{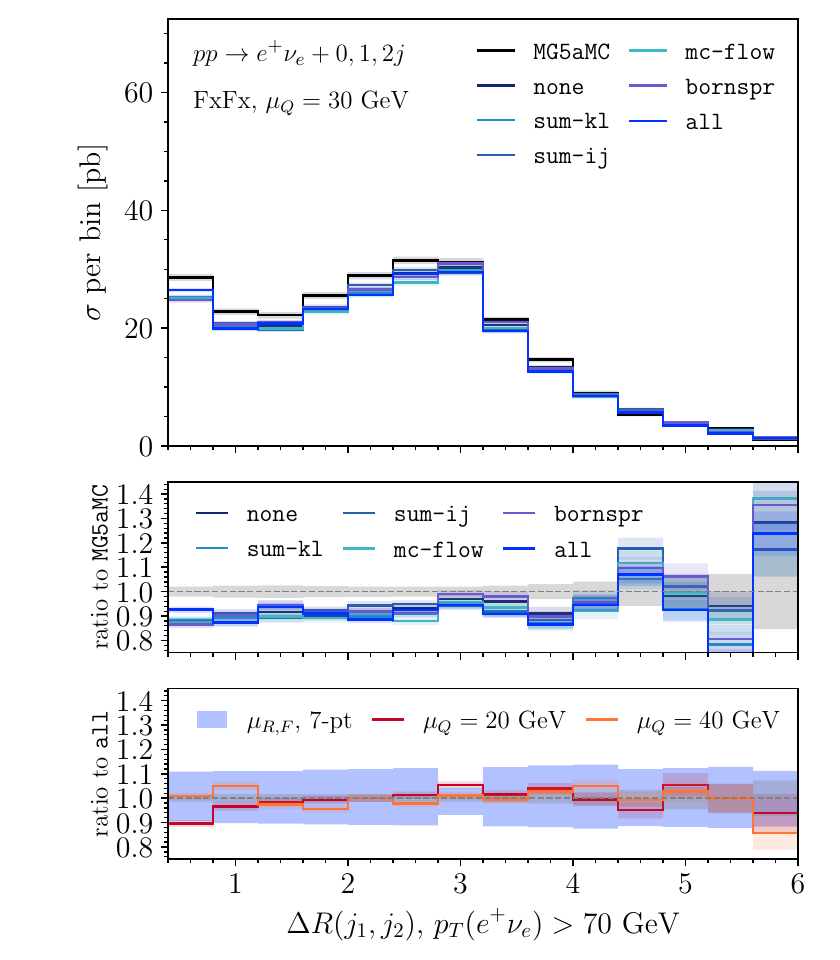}
\caption{FxFx-merged $pp\to e^+\nu_e+0,1,2$ jets at NLO, after the \PYe
shower: transverse momentum of the $e^+\nu_e$ pair (top left), differential
jet rates $\sqrt{d_{01}}$ (top right) and $\sqrt{d_{12}}$ (bottom left),
and separation between the two hardest jets for
$\pT(e^+\nu_e)>70$~GeV (bottom right). Upper panels: cross section per
bin for the variants of Table~\ref{tab:fxfx} with $\mu_Q=30$~GeV. Middle
panels: ratios of the variants of the new implementation to the former
one (\texttt{MG5aMC}). Lower panels: for \vall, ratios to its central
prediction of the envelope of the seven-point variation of the
renormalisation and factorisation scales (band), and of the predictions
with $\mu_Q=20$ and 40~GeV. The bands around the lines show the
statistical uncertainties.}
\label{fig:wjets}
\end{center}
\end{figure*}

In Fig.~\ref{fig:wjets} we show four representative distributions: the
transverse momentum of the $e^+\nu_e$ pair and the differential jet rates
$\sqrt{d_{01}}$ and $\sqrt{d_{12}}$, without cuts, and the separation
$\Delta R(j_1,j_2)$ between the two hardest jets (anti-$k_T$, $R=0.4$,
$\pT>20$~GeV, $|\eta|<4.5$) for $\pT(e^+\nu_e)>70$~GeV. The variants of
the new implementation agree with each other, while they differ from the
former implementation mostly around the merging scale: between 15 and
40~GeV, the new predictions for $\pT(e^+\nu_e)$, $\sqrt{d_{01}}$ and
$\sqrt{d_{12}}$ are lower by up to 12--14\%, whereas they are 2--4\%
higher below 10~GeV and 5--10\% lower at larger values. The
$\Delta R(j_1,j_2)$ distribution is lowered by 5--13\% for
$\Delta R(j_1,j_2)\lesssim3$. These differences stem from two effects: the new
$\clS$-event starting scale $Q=d_{\min}$, Eq.~\eqref{eq:QFxFx}, which
closes the window between the former starting scale and $d_{\min}$
discussed in Section~\ref{sec:old}, in which emissions were suppressed
neither by the FxFx Sudakov factors nor by the shower, and the finite
terms of the Sudakov factors (Section~\ref{sec:newFxFx}), which were
absent in the former implementation. The lower panels show the theoretical
uncertainties of the new predictions, for \vall. The
renormalisation- and factorisation-scale dependence amounts to about
$^{+3}_{-5}\%$ for values of the three inclusive observables below the
merging scale, where the zero-parton sample dominates, and grows to
8--15\% above it. The merging-scale dependence is of a few percent at
most away from the merging-scale value, and it reaches 6\%, 12\% and 12\% for
$\pT(e^+\nu_e)$, $\sqrt{d_{01}}$ and $\sqrt{d_{12}}$, respectively, in
the 20--40~GeV range, similar to the former implementation, for which we
find maximal variations of 5\%, 18\% and 16\%. For $\Delta R(j_1,j_2)$,
it is within about 5\%, except for the smallest separations,
$\Delta R(j_1,j_2)<0.8$, where the prediction with $\mu_Q=20$~GeV is 10\%
lower.

\section{Conclusions and outlook}
\label{sec:conc}
We have revisited the assignment of the shower starting scales in MC@NLO.
By expanding the shower acting on $\clS$ events to first order, we have
derived the conditions under which the damping of the MC counterterms is
consistent with the starting scales, independently of the way in which
the radiation phase space is integrated. Firstly, the starting scale and
the colour flow of $\clS$ events must be generated with Born-level
probability distributions, independently of the radiation variables.
Secondly, the MC counterterms must be damped with the distribution of the
scale that is actually passed to the shower and, separately, cut by its
kinematic limits.

The scale assignment used so far in \aNLOs did not fulfil these
conditions. In particular, the starting scale of $\clS$ events depended
on whether the sampled real-emission configuration lay in a dead zone
of the shower, and on a comparison of its clustering scales with those of
the Born; furthermore, several technical bounds and different reference
scales were used for the damping interval, the shower phase space and the
event file. Besides spoiling, strictly speaking, the NLO accuracy in the
region of intermediate hardness, these features made the predictions
depend on technical aspects of the integration, and prevented the public
release of Born spreading.

We have presented a new construction, mainly for \PYe, which is based on
a single reference scale, the smallest scale in the clustering history
used in FxFx merging, and on a single damped scale per event, from which
the dipole scales are obtained with the kinematic limits of \PYe; the
MC counterterms are their exact first-order expansion. Furthermore, the
real-emission kinematics of initial-state emissions is now constructed
with the same recoil as in \PYe. We have applied
it to ordinary MC@NLO, to FxFx merging, where the starting scale is set
equal to the lowest scale of the Sudakov reweighting, and to \aNLOD,
where the starting scales of the Sudakov factors now coincide with those
used by \PYe for the $\clS$ events, and the stopping scales are computed
natively in \aNLOs. By decoupling the starting scales from the FKS
sectors, the new construction has allowed us to study formulations of the
matching in which the MC counterterms are summed over shower histories
and partitioned like the real emission, in which the FKS sectors are
summed before unweighting, and in which the colour flows are sampled, and
to make Born spreading publicly available, also in conjunction with FxFx
merging.

We have studied the impact of these developments on Drell-Yan, Higgs,
$t\bt$ and $t\bt W^+$ production at the LHC (Section~\ref{sec:res}). When
all of the options are combined, the fraction of negative weights drops
from 6--7\% to below 1\% for Drell-Yan and Higgs production, from 22.6\%
to 8.0\% for $t\bt$ production, and from 15.0\% to 8.1\% for $t\bt W^+$
production, for 3 to 6 times the CPU time. In the transverse momentum of
the hard system, the new construction changes the predictions of the
former implementation by a few percent over most of the Higgs and $t\bt$
distributions, and by up to 12\% and 10\% for Drell-Yan and $t\bt W^+$
production, respectively; the largest effect, around 30~GeV in Drell-Yan
production, arises mainly from the 30~GeV lower bound imposed on $\mu_2$
in the former implementation. The differences between ordinary MC@NLO and
\aNLOD, whose $\clS$ events are now showered under identical conditions,
are of similar size for the four processes, at most about 10\%. Compared with
Ref.~\cite{Frederix:2020trv}, they are much smaller for $t\bt$
production, and somewhat larger for Higgs and Drell-Yan production.

We have studied in detail FxFx-merged $W^+$+jets production with up to
two jets at NLO (Section~\ref{sec:resFxFx}). The event samples are
generated six times faster than with the former implementation, mostly
owing to a faster FxFx clustering. About half of the events pass the
merging veto, and the fraction of negative weights after merging, 14--19\%
(depending on the merging scale) without any option, is considerably
smaller than in the event file, 25\%. Born spreading, which can now be
used with FxFx merging, lowers it by about 3 percentage points without
changing the predictions, and the sums over FKS sectors and over
histories by 1.5 to 2.5 percentage points; with all of these options
combined, it drops to about 9--12\%, for four times the CPU time. About half of
the negative weights stem from $\clH$ events, and two thirds to three
quarters after Born spreading or with all of the options combined, which
motivates the combination
of FxFx merging with \aNLOD discussed below. Physical distributions
change w.r.t.~the former implementation by up to 12--14\% around the
merging scale, while their merging-scale dependence is of similar size.

The new construction may also pave the way for further developments,
of which we mention four. Firstly, the MC counterterms of \aNLOs have
so far been constructed for the global-recoil schemes of \PYe. In
processes with a $t$-channel colour-singlet exchange, such as
vector-boson fusion and $t$-channel single-top production, these schemes
do not respect the colour coherence of the radiation, and a local
(dipole) recoil~\cite{Cabouat:2017rzi} is
preferable~\cite{Jager:2020hkz,Hoche:2021mkv}. Since the MC counterterms
are now evaluated along shower histories, each with its own kinematic
map and the inverse thereof, and since the starting scales and colour
flows are Born-level quantities, the recoil scheme has become an
ingredient of each history, rather than a property of the FKS
parametrisation. This might allow for MC@NLO matching to \PYe with local
recoil~\cite{Frederix:2026loc}.

Secondly, \aNLOD is not available at present in conjunction with FxFx
merging. In FxFx, the short-distance cross sections are reweighted with
Sudakov factors from the hard scale down to the smallest clustering scale
$d_{\min}$, below which the shower takes over
(Section~\ref{sec:newFxFx}), while in \aNLOD the $\Delta$ factor is the
no-emission probability of the shower from the $\clS$-event starting
scales down to the stopping scales of the real emission. With the new
construction, the $\clS$-event starting scale in FxFx coincides with
$d_{\min}$, so that the $\Delta$ factor would naturally continue the
Sudakov suppression of the FxFx reweighting below $d_{\min}$, down to the
hardness of the real emission. Such a combination would be particularly
welcome, since the fraction of negative weights is typically large for
processes with light partons at the Born level (see e.g.~$W^+j$
production in Ref.~\cite{Frederix:2023hom}, and the merged samples of
Section~\ref{sec:resFxFx}). Together with Born spreading, which can
already be used with FxFx (Section~\ref{sec:BS}), it would address both
the negative-weight $\clH$ and $\clS$ events of merged samples.

Thirdly, in MAcNLOPS matching~\cite{Nason:2021xke} the negative-weight
$\clH$ events are removed, and their contribution is compensated by
vetoing, after the first shower emission, a fraction of the $\clS$
events, as has been done for $ZZ$ production~\cite{Che:2026mep}. Since
this compensation relies on the first shower emission off the $\clS$
events, it works best if the MC counterterms describe this emission
accurately, including the starting scales from which it is generated.
This is likely to become more important for processes with more coloured
particles, and thus more colour dipoles, at the Born level. Starting
scales which are consistently shared by the MC counterterms and the
$\clS$ events (and, in \aNLOD, by the $\Delta$ factor), such as those
presented here, should therefore facilitate the application of MAcNLOPS
to such processes.

Finally, the MC counterterms, the starting scales and, for \aNLOD, the
stopping scales and the no-emission probabilities are now all computed
within \aNLOs, without any run-time link to the shower. One could then go
one step further, and generate the first emission off the $\clS$ events
in \aNLOs itself, with the same splitting kernels, kinematic maps, dead
zones and starting scales as those that define the MC counterterms,
while letting the parton shower continue the evolution below the
hardness of this emission, by means of a vetoed (and, for showers not
ordered in transverse momentum, truncated) shower, as is done in
POWHEG~\cite{Nason:2004rx,Frixione:2007vw}. This would potentially lead
to an MC@NLO matching which is agnostic of the parton shower employed for
the subsequent evolution, the dependence on the latter being beyond NLO.
\aNLOD is particularly suited to this purpose: its $\Delta$ factor
already approximates the no-emission probability of the shower between
the $\clS$-event starting scales and the hardness of the first emission,
and its $\clH$ events, which are Sudakov-suppressed in the
soft and collinear regions, are showered from their stopping scales,
i.e.~on the same footing as the $\clS$ events after their first emission.

The code described here is available in the \aNLOs GitHub
repository\footnote{\url{https://github.com/mg5amcnlo/mg5amcnlo}.},
and it will be part of a forthcoming MadGraph7 release.

\section*{Acknowledgements}
We thank Stefano Frixione for a long-standing collaboration on this topic and for useful comments on the manuscript. This work started in 2024,
and its later stages were considerably accelerated by the use of
large-language-model coding assistants, in particular Codex (OpenAI) and Claude (Anthropic).

\bibliographystyle{spphys}
\bibliography{references}

\begin{thebibliography}{10}
\providecommand{\url}[1]{{#1}}
\providecommand{\urlprefix}{URL }
\expandafter\ifx\csname urlstyle\endcsname\relax
  \providecommand{\doi}[1]{DOI \discretionary{}{}{}#1}\else
  \providecommand{\doi}{DOI \discretionary{}{}{}\begingroup
  \urlstyle{rm}\Url}\fi

\bibitem{HEPSoftwareFoundation:2020daq}
T.~Aarrestad, et~al., in \emph{{Snowmass 2021}}, ed. by P.~Canal, et~al.
  (2020).
\newblock \doi{10.5281/zenodo.4009114}.
\newblock {arXiv:2008.13636 [physics.comp-ph]}

\bibitem{CERN-LHCC-2022-005}
{ATLAS Collaboration}, {ATLAS Software and Computing HL-LHC Roadmap}.
\newblock Tech. rep., CERN, Geneva (2022).
\newblock \urlprefix\url{https://cds.cern.ch/record/2802918}

\bibitem{Software:2815292}
{CMS Offline Software and Computing}, {CMS Phase-2 Computing Model: Update
  Document}.
\newblock Tech. rep., CERN, Geneva (2022).
\newblock \urlprefix\url{https://cds.cern.ch/record/2815292}

\bibitem{Frixione:2002ik}
S.~Frixione, B.R. Webber, JHEP \textbf{06}, 029 (2002).
\newblock \doi{10.1088/1126-6708/2002/06/029}

\bibitem{Nason:2004rx}
P.~Nason, JHEP \textbf{11}, 040 (2004).
\newblock \doi{10.1088/1126-6708/2004/11/040}

\bibitem{Frixione:2007vw}
S.~Frixione, P.~Nason, C.~Oleari, JHEP \textbf{11}, 070 (2007).
\newblock \doi{10.1088/1126-6708/2007/11/070}

\bibitem{Jadach:2015mza}
S.~Jadach, W.~P{\l}aczek, S.~Sapeta, A.~Si{\'o}dmok, M.~Skrzypek, JHEP
  \textbf{10}, 052 (2015).
\newblock \doi{10.1007/JHEP10(2015)052}

\bibitem{Nason:2021xke}
P.~Nason, G.P. Salam, JHEP \textbf{01}, 067 (2022).
\newblock \doi{10.1007/JHEP01(2022)067}

\bibitem{vanBeekveld:2025lpz}
M.~van Beekveld, S.~Ferrario~Ravasio, J.~Helliwell, A.~Karlberg, G.P. Salam,
  L.~Scyboz, A.~Soto-Ontoso, G.~Soyez, S.~Zanoli, JHEP \textbf{10}, 038 (2025).
\newblock \doi{10.1007/JHEP10(2025)038}

\bibitem{Nason:2007vt}
P.~Nason, {arXiv:0709.2085 [hep-ph]}  (2007)

\bibitem{Frixione:2007nu}
S.~Frixione, P.~Nason, G.~Ridolfi, {arXiv:0707.3081 [hep-ph]}  (2007)

\bibitem{Alioli:2010xd}
S.~Alioli, P.~Nason, C.~Oleari, E.~Re, JHEP \textbf{06}, 043 (2010).
\newblock \doi{10.1007/JHEP06(2010)043}

\bibitem{Frederix:2020trv}
R.~Frederix, S.~Frixione, S.~Prestel, P.~Torrielli, JHEP \textbf{07}, 238
  (2020).
\newblock \doi{10.1007/JHEP07(2020)238}

\bibitem{Frederix:2023hom}
R.~Frederix, P.~Torrielli, Eur. Phys. J. C \textbf{83}(11), 1051 (2023).
\newblock \doi{10.1140/epjc/s10052-023-12243-x}

\bibitem{Che:2024bsc}
Y.~Che, {Reducing negative weights in MC@NLO by improved implementation of Born
  spreading}.
\newblock {BSc thesis}, Lund University (2024).
\newblock \url{https://lup.lub.lu.se/student-papers/record/9157443}

\bibitem{ElFarkh:2026wtx}
S.~El~Farkh, R.~Frederix, M.~Gouighri, Eur. Phys. J. C \textbf{86}(7), 751
  (2026).
\newblock \doi{10.1140/epjc/s10052-026-16034-y}

\bibitem{Andersen:2020sjs}
J.R. Andersen, C.~G{\"u}tschow, A.~Maier, S.~Prestel, Eur. Phys. J. C
  \textbf{80}(11), 1007 (2020).
\newblock \doi{10.1140/epjc/s10052-020-08548-w}

\bibitem{Andersen:2021mvw}
J.R. Andersen, A.~Maier, Eur. Phys. J. C \textbf{82}(5), 433 (2022).
\newblock \doi{10.1140/epjc/s10052-022-10372-3}

\bibitem{Andersen:2023cku}
J.R. Andersen, A.~Maier, D.~Ma{\^\i}tre, Eur. Phys. J. C \textbf{83}(9), 835
  (2023).
\newblock \doi{10.1140/epjc/s10052-023-11905-0}

\bibitem{Andersen:2024mqh}
J.R. Andersen, A.~Cueto, S.P. Jones, A.~Maier, {arXiv:2411.11651 [hep-ph]}
  (2024)

\bibitem{Alwall:2014hca}
J.~Alwall, R.~Frederix, S.~Frixione, V.~Hirschi, F.~Maltoni, O.~Mattelaer, H.S.
  Shao, T.~Stelzer, P.~Torrielli, M.~Zaro, JHEP \textbf{07}, 079 (2014).
\newblock \doi{10.1007/JHEP07(2014)079}

\bibitem{Frixione:1995ms}
S.~Frixione, Z.~Kunszt, A.~Signer, Nucl. Phys. B \textbf{467}, 399 (1996).
\newblock \doi{10.1016/0550-3213(96)00110-1}

\bibitem{Frixione:1997np}
S.~Frixione, Nucl. Phys. B \textbf{507}, 295 (1997).
\newblock \doi{10.1016/S0550-3213(97)00574-9}

\bibitem{Frederix:2009yq}
R.~Frederix, S.~Frixione, F.~Maltoni, T.~Stelzer, JHEP \textbf{10}, 003 (2009).
\newblock \doi{10.1088/1126-6708/2009/10/003}

\bibitem{Frederix:2012ps}
R.~Frederix, S.~Frixione, JHEP \textbf{12}, 061 (2012).
\newblock \doi{10.1007/JHEP12(2012)061}

\bibitem{Frederix:2021agh}
R.~Frederix, I.~Tsinikos, JHEP \textbf{11}, 029 (2021).
\newblock \doi{10.1007/JHEP11(2021)029}

\bibitem{Bierlich:2022pfr}
C.~Bierlich, et~al., SciPost Phys. Codeb. \textbf{2022}, 8 (2022).
\newblock \doi{10.21468/SciPostPhysCodeb.8}

\bibitem{ATLAS:2024nrd}
G.~Aad, et~al., Eur. Phys. J. C \textbf{84}(10), 1126 (2024).
\newblock \doi{10.1140/epjc/s10052-024-13414-0}

\bibitem{ATLAS:2024mow}
G.~Aad, et~al., Eur. Phys. J. C \textbf{85}(7), 738 (2025).
\newblock \doi{10.1140/epjc/s10052-025-14078-0}

\bibitem{ATLAS:2026wkp}
G.~Aad, et~al., JHEP \textbf{10}, 027 (2026).
\newblock \doi{10.1007/JHEP10(2026)027}

\bibitem{ATLAS:2024moy}
G.~Aad, et~al., JHEP \textbf{05}, 131 (2024).
\newblock \doi{10.1007/JHEP05(2024)131}.
\newblock [Erratum: JHEP 11, 127 (2025)]

\bibitem{CMS:2026fvm}
A.~Gevorgyan, et~al., {arXiv:2609.39653 [hep-ex]}  (2026)

\bibitem{CMS:2026yrb}
A.~Hayrapetyan, et~al., JHEP \textbf{07}, 248 (2026).
\newblock \doi{10.1007/JHEP07(2026)248}

\bibitem{Boos:2001cv}
E.~Boos, et~al., in \emph{{2nd Les Houches Workshop on Physics at TeV
  Colliders}} (2001).
\newblock {arXiv:hep-ph/0109068}

\bibitem{Alwall:2006yp}
J.~Alwall, et~al., Comput. Phys. Commun. \textbf{176}, 300 (2007).
\newblock \doi{10.1016/j.cpc.2006.11.010}

\bibitem{Catani:1993hr}
S.~Catani, Y.L. Dokshitzer, M.H. Seymour, B.R. Webber, Nucl. Phys. B
  \textbf{406}, 187 (1993).
\newblock \doi{10.1016/0550-3213(93)90166-M}

\bibitem{Catani:2001cc}
S.~Catani, F.~Krauss, R.~Kuhn, B.R. Webber, JHEP \textbf{11}, 063 (2001).
\newblock \doi{10.1088/1126-6708/2001/11/063}

\bibitem{Torrielli:2010aw}
P.~Torrielli, S.~Frixione, JHEP \textbf{04}, 110 (2010).
\newblock \doi{10.1007/JHEP04(2010)110}

\bibitem{Hamilton:2012np}
K.~Hamilton, P.~Nason, G.~Zanderighi, JHEP \textbf{10}, 155 (2012).
\newblock \doi{10.1007/JHEP10(2012)155}

\bibitem{Dokshitzer:1978hw}
Y.L. Dokshitzer, D.~Diakonov, S.I. Troian, Phys. Rept. \textbf{58}, 269 (1980).
\newblock \doi{10.1016/0370-1573(80)90043-5}

\bibitem{Catani:1991hj}
S.~Catani, Y.L. Dokshitzer, M.~Olsson, G.~Turnock, B.R. Webber, Phys. Lett. B
  \textbf{269}, 432 (1991).
\newblock \doi{10.1016/0370-2693(91)90196-W}

\bibitem{Catani:1990rr}
S.~Catani, B.R. Webber, G.~Marchesini, Nucl. Phys. B \textbf{349}, 635 (1991).
\newblock \doi{10.1016/0550-3213(91)90390-J}

\bibitem{Lonnblad:2011xx}
L.~Lonnblad, S.~Prestel, JHEP \textbf{03}, 019 (2012).
\newblock \doi{10.1007/JHEP03(2012)019}

\bibitem{Bellm:2015jjp}
J.~Bellm, et~al., Eur. Phys. J. C \textbf{76}(4), 196 (2016).
\newblock \doi{10.1140/epjc/s10052-016-4018-8}

\bibitem{Bewick:2023tfi}
G.~Bewick, et~al., Eur. Phys. J. C \textbf{84}(10), 1053 (2024).
\newblock \doi{10.1140/epjc/s10052-024-13211-9}

\bibitem{Gieseke:2003rz}
S.~Gieseke, P.~Stephens, B.~Webber, JHEP \textbf{12}, 045 (2003).
\newblock \doi{10.1088/1126-6708/2003/12/045}

\bibitem{NNPDF:2021njg}
R.D. Ball, et~al., Eur. Phys. J. C \textbf{82}(5), 428 (2022).
\newblock \doi{10.1140/epjc/s10052-022-10328-7}

\bibitem{Buckley:2014ana}
A.~Buckley, J.~Ferrando, S.~Lloyd, K.~Nordstr{\"o}m, B.~Page, M.~R{\"u}fenacht,
  M.~Sch{\"o}nherr, G.~Watt, Eur. Phys. J. C \textbf{75}, 132 (2015).
\newblock \doi{10.1140/epjc/s10052-015-3318-8}

\bibitem{Cabouat:2017rzi}
B.~Cabouat, T.~Sj{\"o}strand, Eur. Phys. J. C \textbf{78}(3), 226 (2018).
\newblock \doi{10.1140/epjc/s10052-018-5645-z}

\bibitem{Jager:2020hkz}
B.~J{\"a}ger, A.~Karlberg, S.~Pl{\"a}tzer, J.~Scheller, M.~Zaro, Eur. Phys. J.
  C \textbf{80}(8), 756 (2020).
\newblock \doi{10.1140/epjc/s10052-020-8326-7}

\bibitem{Hoche:2021mkv}
S.~H{\"o}che, S.~Mrenna, S.~Payne, C.T. Preuss, P.~Skands, SciPost Phys.
  \textbf{12}(1), 010 (2022).
\newblock \doi{10.21468/SciPostPhys.12.1.010}

\bibitem{Frederix:2026loc}
R.~Frederix, P.~Torrielli (2026).
\newblock In preparation

\bibitem{Che:2026mep}
Y.~Che, R.~Frederix, Eur. Phys. J. C \textbf{86}(7), 870 (2026).
\newblock \doi{10.1140/epjc/s10052-026-16152-7}

\end{thebibliography}

\end{document}